\documentclass[aps,prl,reprint,showpacs,superscriptaddress,longbibliography]{revtex4-2}
\usepackage{mathtools,amssymb,graphicx,units}

\usepackage[plainpages=false,pdfpagelabels,colorlinks=true,linkcolor=red,urlcolor=blue,citecolor=blue,pdftitle={Title},pdfauthor={},pdfdisplaydoctitle=true,pdfduplex=DuplexFlipLongEdge]{hyperref}
\usepackage{verbatim}
\usepackage[english]{babel}
\usepackage[dvipsnames]{xcolor}

\usepackage[normalem]{ulem}

\newcommand{\beginsupplement}{%
        \setcounter{table}{0}
        \renewcommand{\thetable}{S\arabic{table}}%
        \setcounter{figure}{0}
        \renewcommand{\thefigure}{S\arabic{figure}}%
}

\begin{document}

\title{Quantum Critical Points and the Sign Problem}
\author{R. Mondaini}
\email{rmondaini@csrc.ac.cn}
\author{S. Tarat}
\email{tarats@csrc.ac.cn}
\affiliation{Beijing Computational Science Research Center, Beijing 100193, China}
\author{R.T. Scalettar}
\email{scalettar@physics.ucdavis.edu}
\affiliation{Department of Physics, University of California,
Davis, CA 95616, USA}
 
\begin{abstract}
The ``sign problem'' (SP) is the fundamental limitation to simulations of strongly correlated materials in condensed matter physics, solving quantum chromodynamics at finite baryon density, and computational studies of nuclear matter.  As a result, it is part of the reason fields such as ultra-cold atomic physics are so exciting: they can provide quantum emulators of models that could not otherwise be solved, due to the SP. For the same reason, it is also one of the primary motivations behind quantum computation. It is often argued that the SP is not intrinsic to the physics of particular Hamiltonians, since the details of how it onsets, and its eventual occurrence, can be altered by the choice of  algorithm or many-particle basis. Despite that, we show that the SP in determinant quantum Monte Carlo (DQMC) is quantitatively linked to quantum critical behavior. We demonstrate this via simulations of a number of fundamental models of condensed matter physics, including the spinful and spinless Hubbard Hamiltonians on a honeycomb lattice and the ionic Hubbard Hamiltonian, all of whose critical properties are relatively well understood. We then propose a reinterpretation of the low average sign for the Hubbard model on the square lattice when away from half-filling, an important open problem in condensed matter physics, in terms of the onset of pseudogap behavior and exotic superconductivity. Our study charts a path for exploiting the average sign in QMC simulations to understand quantum critical behavior, rather than solely as an obstacle that prevents quantum simulations of many-body Hamiltonians at low temperature.
 
\end{abstract}

\maketitle

%%%%%%%%%%%%%%%%%%%%%%%%%%%%%%%%%%%%%%%%%%%%%%%%%%%%%%%%%%%%%%%%%%
%%%%%%%%%%%%%%%%%%%%%%%%%%%%%%%%%%%%%%%%%%%%%%%%%%%%%%%%%%%%%%%%%%
% \vskip0.10in 
% {\bf \begin{center} I.  Introduction \end{center}}  \label{sec:Introduction}
%%%%%%%%%%%%%%%%%%%%%%%%%%%%%%%%%%%%%%%%%%%%%%%%%%%%%%%%%%%%%%%%%%
%%%%%%%%%%%%%%%%%%%%%%%%%%%%%%%%%%%%%%%%%%%%%%%%%%%%%%%%%%%%%%%%%%
Over the last several decades, quantum Monte Carlo (QMC) simulations have provided great insight into challenging strong correlation problems in chemistry~\cite{hammond94,needs20}, condensed matter~\cite{ceperley95,foulkes01}, nuclear~\cite{carlson15}, and high energy physics~\cite{degrand06}. In all these areas, however, the  sign problem (SP), which occurs when the probability for specific quantum configurations in the importance sampling becomes negative, significantly constrains their application. Solving, or at least mitigating, the SP is one of the central endeavors of computational physics. The extent and importance of the effort is indicated by the many proposed solutions, and their continued development over the last three decades --- see the Supplemental Materials (SM) for a review~\cite{SM}.

Despite enormous effort, the SP remains unsolved. In fact, the lack of progress is one of the main driving forces behind a number of large-scale efforts, including the quest for quantum emulators~\cite{esslinger10,bloch12,schafer20} as well as quantum computing itself~\cite{preskill18,clemente20}. One of the most fundamental mysteries concerns the possible link between the sign problem and the underlying physics of the Hamiltonian being investigated.

Here, instead of challenging this NP-hard problem~\cite{troyer05}, or proposing solutions that can partially ameliorate its behavior~\cite{Hangleiter2020,Wan2020}
we show that there is a clear \textit{quantitative} connection between the behavior of the average sign $\langle {\cal S} \rangle$ in the widely used Determinant Quantum Monte Carlo (DQMC) method and several quantum phase transitions: that of the semimetal to antiferromagnetic Mott insulator (AFMI) of Dirac fermions in the  spinful [SU(2)] honeycomb-Hubbard Hamiltonian~\cite{meng10,sorella12}, 
the band to correlated insulator transition 
~\cite{fabrizio99,craco08,garg14}, and charge density wave transitions of spinless [U(1)] fermions on a honeycomb lattice~\cite{huffman14,Wang2014}. In the first case, simulations at half-filling, where the quantum critical point (QCP) occurs, are SP free. We introduce a small doping $\mu$ and show, in the limit $\mu \to 0^+$ at temperature $T\to0$, that $\langle {\cal S} \rangle$ evolves rapidly as we tune through the QCP.

Our second illustration, the ionic Hubbard model, has a SP even at half-filling. Here the average sign undergoes an abrupt drop at the band-insulator (BI) to correlated metal (CM) transition. The third case, spinless fermions on a honeycomb lattice, also has a semimetal to (charge) insulator transition. Its interest relies on the existence of a SP-free approach. Studying it with a method which contains an `unnecessary' SP lends insight into the key question of the algorithm dependence of links between the SP and the physics of model Hamiltonians.

These three discussions establish a link between known physics of the models and the fermion sign. Having made that connection, we then turn to the iconic square lattice Hubbard model whose physics has {\it not} been conclusively established. We find that the {\it onset} of the SP occurs in a dome-shaped region of the filling-temperature phase space underneath that of the pseudogap (PG) physics. The SP is well-enough controlled in the PG phase to obtain reliable results for various observables, including the pairing correlations in various channels, exhibiting dominant enhancement for $d$-wave symmetry. Because it onsets exponentially in inverse temperature, the SP provides a rather sharp demarcation of the regime, which mimics the superconducting dome of the cuprates~\cite{Keimer2015}. Although the SP prevents DQMC from resolving a signal of a $d$-wave transition, the ground work established for the honeycomb lattice and BI-CM models suggests that this sign problem dome might be linked to the onset of a superconducting phase. In the conclusions we will make connections which will further corroborate the generality of our results to other models and QMC methods.

%%%%%%%%%%%%%%%%%%%%%%%%%%%%%%%%%%%%%%%%%%%%%%%%%%%%%%%%%%%%%%%%%%
%%%%%%%%%%%%%%%%%%%%%%%%%%%%%%%%%%%%%%%%%%%%%%%%%%%%%%%%%%%%%%%%%%
\vskip0.10in 
{\bf \begin{center} The Sign Problem; Model and Methodology \end{center}}  \label{sec:ModelandMethods}
%%%%%%%%%%%%%%%%%%%%%%%%%%%%%%%%%%%%%%%%%%%%%%%%%%%%%%%%%%%%%%%%%%
%%%%%%%%%%%%%%%%%%%%%%%%%%%%%%%%%%%%%%%%%%%%%%%%%%%%%%%%%%%%%%%%%%

The origin of the sign problem can be understood in two related classes of algorithms, `world-line' Quantum Monte Carlo (WLQMC)~\cite{hirsch82} and Green's function Quantum Monte Carlo (GFQMC)~\cite{ceperley81,lee92}, by considering Feynman's path integral approach, which provides a mapping of quantum statistical mechanics in $D$ dimensions to classical statistical mechanics in $D+1$ dimensions.
Paralleling Feynman's original exposition for the operator $e^{-{\rm i} \hat H t /\hbar}$, the imaginary time evolution operator $e^{-\beta \hat H}$ is subdivided into $L_\tau$ incremental pieces $\hat U_{\Delta \tau} = e^{-\Delta \tau \hat H}$, where $L_\tau \, \Delta \tau = \beta$, the inverse temperature.
%% Following Feynman's construction, 
Complete sets of states $\hat {\cal I}_\tau = \sum_{S_{\tau}} | S_{\tau} \rangle \langle S_{\tau} |$ are introduced between each $\hat U_{\Delta \tau}$ so that the partition function ${\cal Z} = {\rm Tr} \, e^{-\beta \hat H}$  becomes a sum over the {\it classical} degrees of freedom associated with the spatial labels of each $\hat {\cal I}_\tau$, and also an additional imaginary time index denoting the location $\tau = 1,2, \ldots, L_\tau$ of $\hat {\cal I}_\tau$ in the string of operators  $\hat U_{\Delta \tau}$. The quantity being summed in the calculation of ${\cal Z}$ is the product of matrix elements $\langle S_{\tau} | \hat U_{\Delta \tau} | S_{\tau+1}\rangle $.

In such WLQMC/GFMC methods, the SP arises when $\prod_{\tau} \langle S_{\tau} | \hat U_{\Delta \tau} | S_{\tau+1}\rangle < 0$. Negative matrix elements are unavoidable for itinerant fermionic models in $D>1$ because their sign depends on the number of fermions intervening between two particles undergoing exchange, and hence changes as the particle positions are updated. The basis dependence of the sign problem is apparent by considering intermediate states $|S_\tau \rangle$ chosen to be eigenstates $|\phi_{\alpha} \rangle$ of $\hat H$.  In that case the matrix elements are just $e^{-\Delta \tau E_\alpha}$ and hence are trivially positive definite.  Of course, since the eigenstates of $\hat H$ are unknown, this is not a practical choice in any non-trivial situation. Moreover, the SP can generally be avoided for bosonic or spin models as long as the lattice is bipartite. Nonetheless, even bosonic and spin Hamiltonians can have negative matrix elements on frustrated geometries~\cite{henelius00}, especially for antiferromagnetic models, emphasizing the SP is not solely a consequence of Fermi statistics.

`Auxiliary field' Quantum Monte Carlo (AFQMC) algorithms~\cite{blankenbecler81,white89,zhang97} typically have a much less severe SP than WLQMC~\cite{Iazzi2016,SM}. They are based on the observation that the trace of an exponential of a quadratic form of fermionic operators can be done analytically, resulting in the determinant of a matrix of dimension set by the cardinality of fermionic operators. The determinant is the product $\prod_j \big( 1 + e^{-\beta \varepsilon_j}\big)$, where $\varepsilon_j$ are the non-interacting energy levels, and is always positive.

\begin{figure*}[htbp]
\includegraphics[width=2\columnwidth]{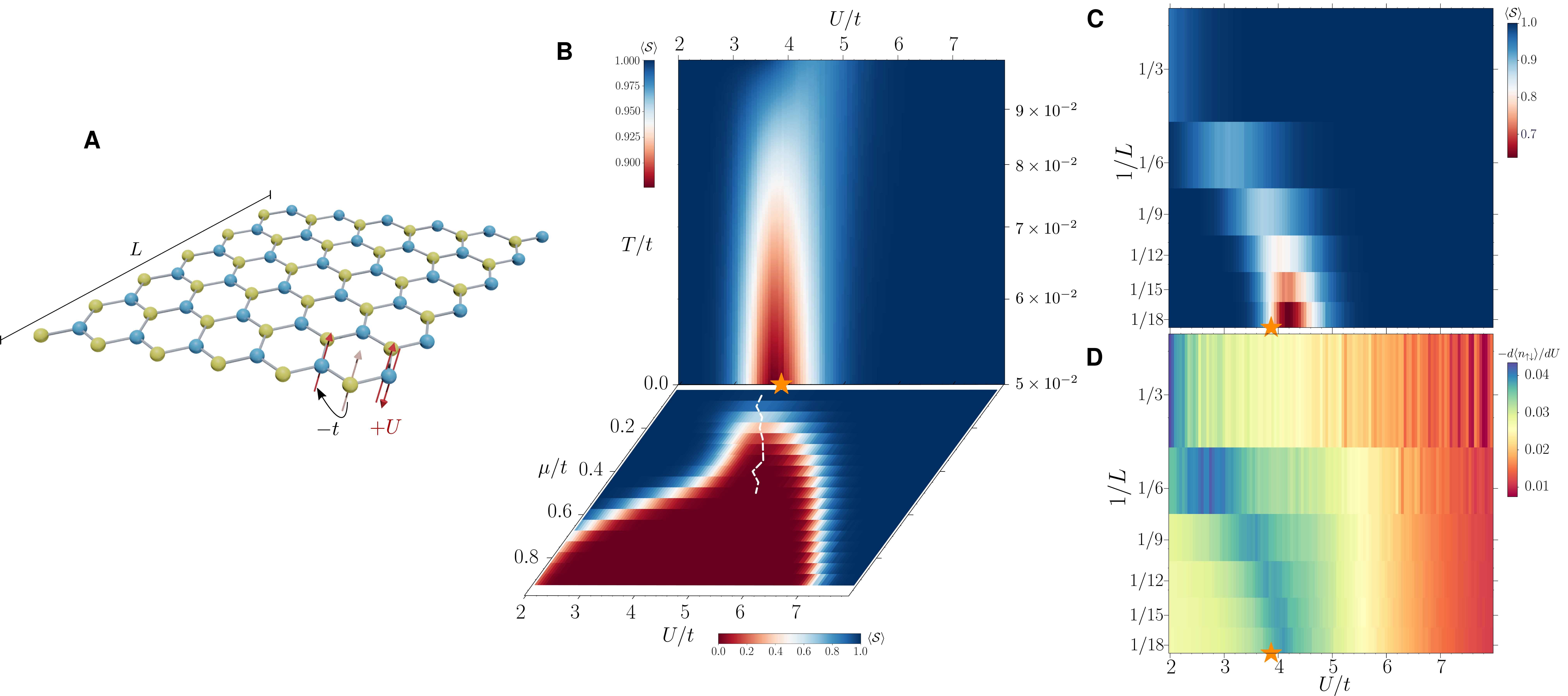}
% \vspace{-0.6cm}
\caption{\textbf{The SU(2) Hubbard model on the honeycomb lattice.} (\textbf{A}) Cartoon depicting a honeycomb lattice with $N=2L^2$ sites ($L=6$ here), accompanied by the relevant terms in $\hat H$. (\textbf{B}) Contour plot of the average $\langle {\cal S}\rangle$ in the $T/t$ ($\mu/t$) vs $U/t$ in the upper (lower) panel. Here $L = 9$ and $\mu/t=0.1$ ($T/t = 1/20$) in the upper (lower) panel. (\textbf{C}) The average sign extrapolated with the linear system size $L$, using $T/t = 1/20$ and $\mu/t = 0.1$. (\textbf{D}) Similar extrapolation as in \textbf{C}, but displaying a local quantity (the derivative of the double occupancy), which is an indicator of the QCP. In all panels with data, the prediction for the ground-state phase transition occurring at $U_c/t = 3.869$~\cite{sorella12} is depicted by a star marker.}
\label{fig:1}
\end{figure*}

If interactions are present,
quartic terms in $\hat H$ are reduced to quadratic ones with a Hubbard-Stratonovich (HS) transformation. The trace of the resulting product of exponentials of quadratic forms can be performed, but now they each depend on a different, i.e.~imaginary-time dependent, auxiliary field.
The resulting determinant is no longer guaranteed to be positive; the consequence is the SP since the HS field needs to be sampled stochastically in order to compute operator expectation values.  

In AFQMC, the trace over fermionic degrees of freedom is done for all species (i.e.~all spin and orbital indices $\alpha$). If there is no hybridization between different $\alpha$, each trace gives an individual determinant.  In some situations, particle-hole, time-reversal, or other symmetries~\cite{loh90,wu05,li16} impose a relation between the determinants for different $\alpha$'s, and as a consequence the negative determinants always come in pairs. Low temperature (ground-state) properties can be accessed in such `sign problem free' cases, and a host of interesting quantum phase transitions has been explored~\cite{chandrasekharan10,berg12,wang15,li19}. If such a partnering does not occur, a reasonable rule of thumb is that the average sign $\langle {\cal S} \rangle$ is sufficiently bounded away from zero with measurements that exhibit sufficiently small error bars for $T \gtrsim W/20\,$-$\,W/40$, at intermediate interaction strengths (of the same order as the bandwidth $W$)~\footnote{This is just a rough guideline; the precise onset of the SP is determined by lattice geometry, doping (chemical potential), and interaction strength. A catalog of the SP in DQMC for the single band Hubbard model in different situations is given in Ref.~\cite{iglovikov15}.}.

The DQMC methodology~\cite{blankenbecler81,white89} we use is a specific implementation of AFQMC. We employ the discrete HS transformation introduced by Hirsch~\cite{hirsch83} and choose the Trotter discretization $\Delta \tau$ such that systematic errors in $\langle {\cal S} \rangle$ and other observables are of the same order as statistical sampling errors. (See~\cite{SM} for additional details.)

We mostly consider models in which two (spin) species of itinerant electrons hop on a lattice with an on-site repulsion, i.e., variants of the Hubbard Hamiltonian,
\begin{align}
\hat H = 
%% &- \sum_{\langle ij \rangle \,\sigma} t_{ij}
&- \sum_{ij \,\sigma}  t_{ij}
\big( \, \hat c^{\dagger}_{i\sigma} \hat c^{\phantom{\dagger}}_{j\sigma}
+\hat c^{\dagger}_{j\sigma} \hat c^{\phantom{\dagger}}_{i\sigma} \, \big)
- \sum_{i \sigma} \mu_i \, \hat n_{i\sigma} 
\nonumber \\
&+ U \sum_{i} \left(\hat n_{i\uparrow} -\frac{1}{2} \right) 
\, \left(\hat n_{i\downarrow} - \frac{1}{2}\right).
\label{eq:ham_spinful}
\end{align}
Here $\hat c^{\dagger}_{j\sigma} (\hat c^{\phantom{\dagger}}_{j\sigma} )$ are creation (destruction) operators at site $i$ with spin $\sigma$ and $\hat n_{i\sigma} = \hat c^{\dagger}_{i\sigma} \hat c^{\phantom{\dagger}}_{i\sigma}$ is the number operator. In Sec.~I, $i,j$ are near-neighbor sites on a honeycomb lattice, with $t_{ij}=t$. As a consequence of particle-hole symmetry (PHS), $\mu_i=0$ corresponds to half-filling, and $\rho = \langle \hat n_{i\sigma} \rangle = 1/2$, for arbitrary $U$ and temperature $T$.
In Sec.~II, we consider a $t_{ij}=t$ square lattice with $\mu_i = +\Delta$ on one sublattice and $\mu_i = -\Delta$ on the other, a situation which has a SP even at half-filling, but which is mild enough to allow its phase diagram to be established with reasonable reliability. Sec.~III concerns a single species model with interactions between fermions on neighboring sites, notable because a SP-free QMC formulation exists~\cite{Wang2014,ZiXiang2015}.

All these models have QCPs which have been located to fairly high precision and so serve as testbeds for demonstrating that the average sign can be used as an alternate means to study the onset of quantum criticality. In our final investigation, Sec.~IV, we consider the doped, spinful, square lattice Hubbard model, much of whose low temperature physics remains shrouded in mystery. We correlate the behavior of the SP with some of its properties at intermediate temperature, and then describe what might be inferred concerning one the most elusive puzzles--the presence of a low temperature superconducting dome.

%%%%%%%%%%%%%%%%%%%%%%%%%%%%%%%%%%%%%%%%%%%%%%%%%%%%%%%%%%%%%%%%%%
\vskip0.10in 
{\bf \begin{center} I. Semimetal to AFMI on a Honeycomb Lattice\end{center}} \label{sec:Honeycomb}
%%%%%%%%%%%%%%%%%%%%%%%%%%%%%%%%%%%%%%%%%%%%%%%%%%%%%%%%%%%%%%%%%%

On a honeycomb lattice (Fig.~\ref{fig:1}\textbf{A}), the $U=0$ Hubbard Hamiltonian has a semi-metallic density of states which vanishes linearly at $E=0$. Its dispersion relation $E({\bf k})$ has Dirac points in the vicinity of which the kinetic energy varies linearly with momentum. Unlike the square lattice that displays AF order for all $U \neq 0$, the honeycomb Hubbard model at $T\to0$ remains a semimetal for small nonzero $U$, turning to an AF insulator only for $U$ exceeding a critical $U_c$. Early DQMC and series expansion calculations estimated $U_c \sim 4\,t$~\cite{paiva05}, with subsequent studies~\cite{meng10,sorella12} 
%% narrowing it closer to $U_c/t = 3.869$.
yielding the more precise value $U_c/t = 3.869$.

The upper panel of Fig.~\ref{fig:1}\textbf{B} gives $\langle {\cal S} \rangle$ in the $U$-$T$ plane. By introducing a small non-zero $\mu=0.1$, we can induce a SP which begins to develop at $T/t \sim 0.1$.  As $T$ is lowered further the average sign deviates from $\langle {\cal S} \rangle =1$ in a relatively narrow window of $U/t$ close to the known $U_c$. In turn, we show the $\langle {\cal S} \rangle$ on the $U-\mu$ plane at fixed $T/t=0.05$ in the lower panel of Fig.~\ref{fig:1}\textbf{B}. For large $\mu$ the sign is small for a broad swath of interaction values. As $\mu$ decreases this region pinches down until it terminates close to $U_c$; the dashed white line displays the minimum $\langle {\cal S} \rangle$ in the relevant range. In both panels, the behavior of the average sign outlines the quantum critical fan that extends above the QCP.

Figure~\ref{fig:1}\textbf{C} shows a finite size extrapolation of $\langle {\cal S} \rangle$ in the $1/L-U$ plane, where $L$ is the linear lattice size. Just as $\langle {\cal S} \rangle$ worsens with increasing $\beta$, it is also known to deviate increasingly from $\langle {\cal S} \rangle=1$ with growing $L$~\cite{white89}. What these data further indicate is that the extrapolation $L\rightarrow \infty$ clearly reveals $U_c$ in the presence of a small chemical potential. So far, we have exclusively used $\langle {\cal S} \rangle$ in locating $U_c$. Original investigations employed more `traditional' (and more physical) correlation functions such as the AF structure factor and conductivity. For comparison to the evolution of $\langle {\cal S} \rangle$, Fig.~\ref{fig:1}\textbf{D} shows one example, the rate of change of the double occupancy $D$, again in the $1/L-U$ plane. A peak in $-dD/dU$ indicates where local moments $\langle m^2 \rangle$ are growing most rapidly. The similarity between Figs.~\ref{fig:1}\textbf{C} and~\ref{fig:1}\textbf{D} emphasizes how $\langle {\cal S} \rangle$ is tracking the physics of the model in a way remarkably similar to $\langle m^2 \rangle$. The combination of the three limits, $\mu, \beta, L$, unequivocally points out the QCP location; the SM~\cite{SM} contains further discussion, and other observables.

\begin{figure}[htbp]
\includegraphics[width=1\columnwidth]{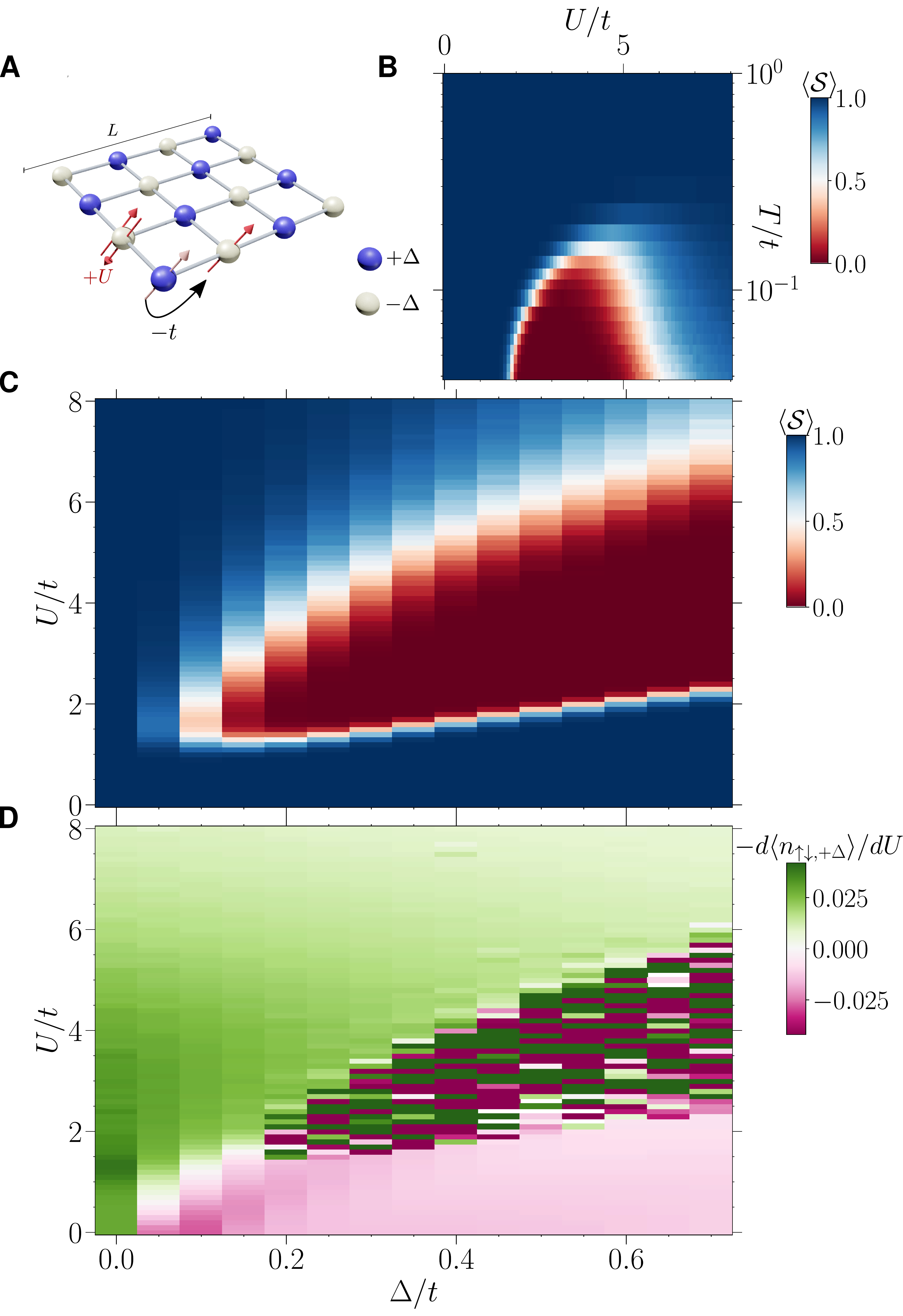}
% \vspace{-0.6cm}
\caption{\textbf{The SU(2) Ionic Hubbard model on the square lattice.} (\textbf{A}) Cartoon depicting a square lattice with $N=L^2$ sites ($L=4$ here), accompanied by the relevant terms in $\hat H$. (\textbf{B}) Contour plot of the average $\langle {\cal S}\rangle$ in the $U/t$ vs. $T/t$ plane, with staggered potential $\Delta/t = 0.5$. (\textbf{C}) The contour of $\langle {\cal S}\rangle$ as function of the competing parameters $U/t$ and $\Delta/t$, at a temperature $T/t = 1/24$. (\textbf{D}) The corresponding derivative of the double occupancy on the $+\Delta$ sites at the same parameters as in \textbf{C}. In all data, Trotter discretization is chosen as $\Delta\tau = 0.1$ and the lattice size is $L=12$.}
\label{fig:2}
\end{figure}

\begin{figure}[htbp]
\includegraphics[width=1\columnwidth]{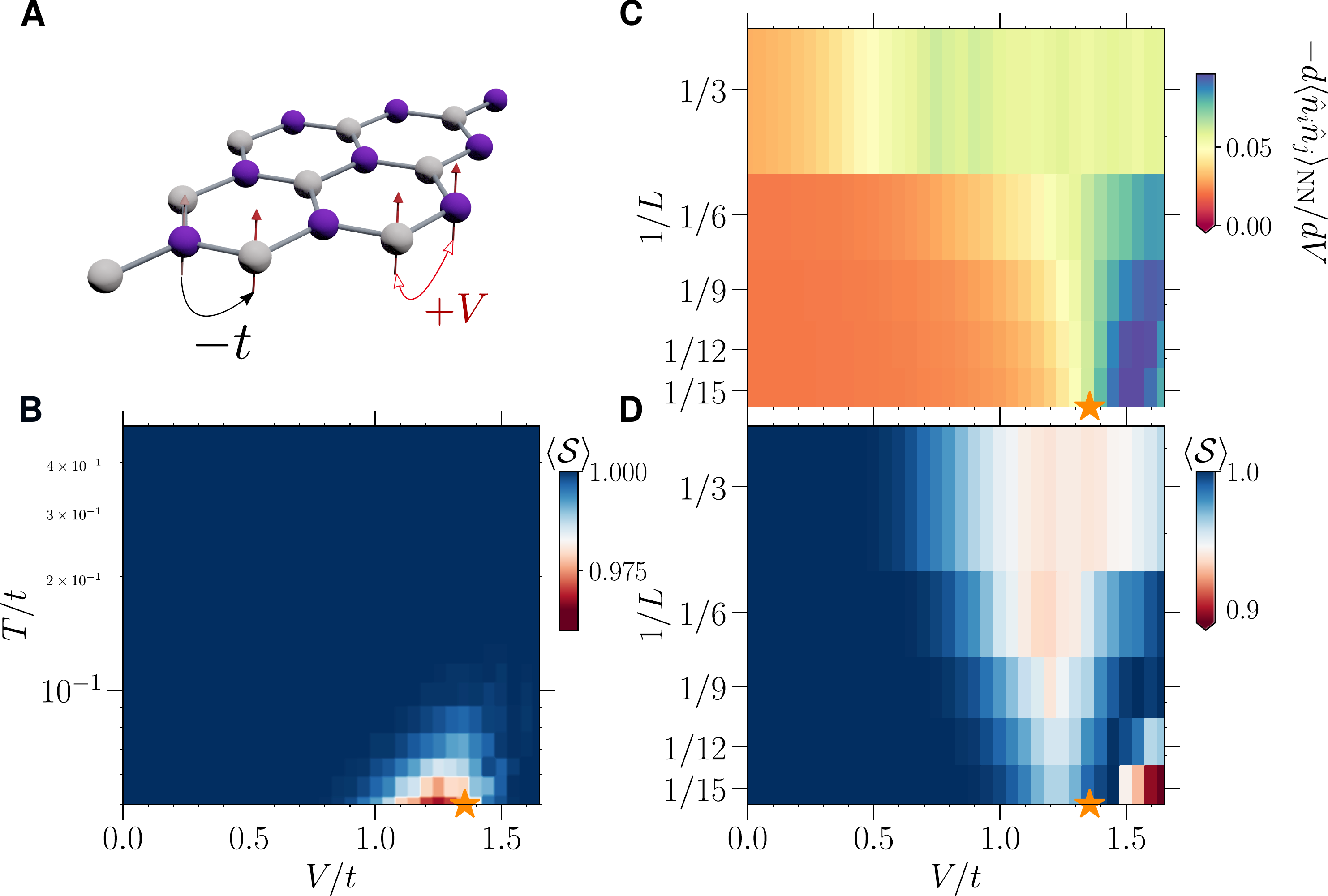}
% \vspace{-0.6cm}
\caption{\textbf{The U(1) Hubbard model on the honeycomb lattice.} (\textbf{A}) Schematics of the spinless fermion Hamiltonian [Eq.~\eqref{eq:ham_spinless}] with nearest-neighbor interaction on a lattice with $L=3$. Temperature extrapolation of the average $\langle {\cal S}\rangle$ as function of the nearest-neighbor interaction $V/t$, for a lattice with $L=9$. (\textbf{D}) [(\textbf{C})] Extrapolation of the average sign [derivative of the nearest-neighbor correlation in respect to $V$]  with the inverse of the linear size $L$ along a range of interactions, at a temperature $T$ that scales with the system size $T/t = 0.0375/ L\Delta\tau$. Here $\langle {\cal S}\rangle$ marginally increases when tackling larger sizes, indicating that the dynamical critical exponent $z$ in the scaling with $L_\tau/L^z$ is larger than one~\cite{Rieger1994,SM}; we used $z=1$ above. In all data, Trotter discretization is chosen as $\Delta\tau = 0.1$. As for the SU(2) case, the star marker depicts the best known value of the interactions that trigger the Mott insulating phase; here with CDW order at the ground-state~\cite{ZiXiang2015}.}
\label{fig:3}
\end{figure}

%%%%%%%%%%%%%%%%%%%%%%%%%%%%%%%%%%%%%%%%%%%%%%%%%%%%%%%%%%%%%%%%%%
%%%%%%%%%%%%%%%%%%%%%%%%%%%%%%%%%%%%%%%%%%%%%%%%%%%%%%%%%%%%%%%%%%
\vskip0.10in 
{\bf \begin{center} II.  Ionic Hubbard BI to AF Transition\end{center}} \label{sec:BIAF}
%%%%%%%%%%%%%%%%%%%%%%%%%%%%%%%%%%%%%%%%%%%%%%%%%%%%%%%%%%%%%%%%%%
%%%%%%%%%%%%%%%%%%%%%%%%%%%%%%%%%%%%%%%%%%%%%%%%%%%%%%%%%%%%%%%%%%

Among the different types of non-conducting states are `band insulators' (BI), in which the chemical potential lies in a gap in the non-interacting density of states (DOS), and `Mott insulators' (MI) in which strong repulsive interactions prevent hopping at commensurate filling. The evolution from BI to MI is a fascinating issue in condensed matter 
physics~\cite{fabrizio99,kampf03,garg06,paris07,craco08,garg14}.
In the ionic Hubbard model we investigate here, a staggered site energy $\mu_i = \pm \Delta$ on the two sublattices of a square lattice (Fig.~\ref{fig:2}\textbf{A}) leads to a dispersion relation $E(k) = \pm \sqrt{\varepsilon(k)^2 + \Delta^2}$ with $\varepsilon(k) = -2t\,({\rm cos}k_x + {\rm cos} k_y\,)$. The resulting DOS vanishes in the range $-\Delta < E< +\Delta$ in which the lattice is half-filled, resulting in a BI. The occupation of the `low energy' sites $\mu_i = -\Delta$ is greater than that of the `high energy' sites $\mu_i=+\Delta$, so that there is a trivial charge density wave (CDW) order associated with an explicit breaking of the sublattice symmetry in the Hamiltonian.

An onsite repulsion $U$ disfavors this density modulation: The potential energy $U n_{i\uparrow} n_{i\downarrow}$ is higher than for a uniform occupation. Thus the driving physics of the BI, the staggered site energy $\Delta$ and that of the MI, the repulsion $U$, are in competition. Although the simplest scenario is a direct BI to MI transition with increasing $U$, one of the more exotic possible outcomes is the emergence of a metallic phase when these two energy scales are in balance and neither type of insulator can dominate the behavior. Past DQMC simulations suggest this less trivial case occurs, and have used the temperature dependence of the dc conductivity to bound the metallic phase~\cite{paris07,chattopadhyay19}.

Here we investigate how this physics might be reflected in the average sign. Figure ~\ref{fig:2}$\,$\textbf{B} shows $\langle {\cal S} \rangle$ in the $U/t-T/t$ plane at $\Delta=0.5\,t$.  As $T$ is lowered, $\langle {\cal S} \rangle$ deviates from unity for a range of intermediate $U$ values. Figure~\ref{fig:2}$\,$\textbf{C} gives the behavior in the $U/t-\Delta/t$ plane at fixed low $T=t/24$. The central result is that $\langle {\cal S} \rangle$ is small in a region which maps well with the previously determined boundaries of the metallic phase~\cite{paris07,chattopadhyay19}. This is emphasized by a comparison to Fig.~\ref{fig:2}$\,$\textbf{D}, which uses one of the `traditional' methods for phase boundary location, namely the behavior of the double occupancy. 
The BI has a low occupancy, and hence very low double occupancy on the $+\Delta$-sites. Increasing $U$ smooths out the density, so that the double occupancy on the $+\Delta$-sites {\it increases}: $d\langle n_{\uparrow\downarrow,+\Delta}\rangle/dU>0$. In contrast, in the MI region, $U \gtrsim \Delta$, the physics is that of the usual Hubbard Hamiltonian and double occupancy {\it decreases} as $U$ grows:
$d\langle n_{\uparrow\downarrow,+\Delta}\rangle/dU<0$.

In the CM region between BI and MI, however, obtaining a relevant signal-to-noise ratio for the traditional observables is exponentially challenging precisely because the average sign vanishes in this region.
The `phase diagram' obtained by $\langle{\cal S}\rangle$ (Fig.~\ref{fig:2}$\,$\textbf{C}) is remarkably similar to that given by the physical observable, the rate of change of double occupancy with $U$ (Fig.~\ref{fig:2}$\,$\textbf{D}) 

As in the determination of the QCP for the spinful Hubbard model on a honeycomb lattice, $\langle {\cal S} \rangle$ emerges as more than a mere nuisance, but as a harbinger of the physics. An in-depth similarity between these two situations is discussed in the SM~\cite{SM}, where we show that the BI-metal QCP is again uniquely identified by the $1/L$ scaling of $\langle {\cal S} \rangle$, in precise analogy with the honeycomb case. These results suggest the existence of a quantum critical region associated with the CM phase and the vanishing $\langle {\cal S} \rangle$.

%%%%%%%%%%%%%%%%%%%%%%%%%%%%%%%%%%%%%%%%%%%%%%%%%%%%%%%%%%%%%%%%%%
%%%%%%%%%%%%%%%%%%%%%%%%%%%%%%%%%%%%%%%%%%%%%%%%%%%%%%%%%%%%%%%%%%
\vskip0.10in 
{\bf \begin{center} III.  An `Unnecessary' Sign Problem\end{center}}  \label{sec:spinlessfermions}
%%%%%%%%%%%%%%%%%%%%%%%%%%%%%%%%%%%%%%%%%%%%%%%%%%%%%%%%%%%%%%%%%%
%%%%%%%%%%%%%%%%%%%%%%%%%%%%%%%%%%%%%%%%%%%%%%%%%%%%%%%%%%%%%%%%%%

We now consider spinless fermions, where the on-site Hubbard interaction $U$, made irrelevant by the Pauli principle, is replaced by an intersite repulsion $V$,
\begin{align}
\hat H = 
&- t\sum_{\langle ij \rangle } 
\big( \, \hat c^{\dagger}_{i} \hat c^{\phantom{\dagger}}_{j}
+ \hat c^{\dagger}_{j} \hat c^{\phantom{\dagger}}_{i} \, \big)
+ V \sum_{\langle ij \rangle } 
\hat n_{i}  \hat n_{j}.
%% \big( \, n_{i} -\frac{1}{2} \, \big) 
%% \, \big( \, n_{j} - \frac{1}{2} \, \big)
\label{eq:ham_spinless}
\end{align}
 
Equation~\ref{eq:ham_spinless} provides an example of a model where the SP can be completely solved by utilizing special techniques such as the fermion bag in the Continuous Time QMC approach~\cite{chandrasekharan10}, or by going to a different basis employing a Majorana representation of the fermions in the AFQMC method~\cite{ZiXiang2015}, as long as the system is on a bipartite lattice and $V>0$. The standard Blankenbecler, Scalapino, and Sugar (BSS) approach ~\cite{blankenbecler81}, on the other hand, manifestly displays a SP in the low temperature regime. Nevertheless, in order to study the sign and its connection with the underlying physics, we use a BSS based algorithm to investigate the system on a honeycomb lattice (Fig.~\ref{fig:3}$\,$\textbf{A}). Consideration of this `unnecessary' SP allows us to address fundamental issues related to the algorithm dependence of links between the SP and the physics of model Hamiltonians.

At $T=0$, the model displays a QPT between a Dirac semimetal and an insulating staggered CDW state as the interaction is tuned through a critical value $V_{c}$~\cite{Wang2014}. At large $V$, the repulsive interaction favours a CDW state, distinguished from that of the ionic Hubbard model by the fact that there is no staggered external field here -- the CDW phase is a result of {\it spontaneous} symmetry breaking. As $V$ is reduced, increasing quantum fluctuations due to hopping finally destroy the CDW state, resulting in a Dirac semimetal for $V<V_{c}$. Accurate estimates based on SP free methods yield $V_{c} \sim 1.35t$~\cite{ZiXiang2015}.

In Fig.~\ref{fig:3}$\,$\textbf{B}, we show a map of the temperature extrapolation of $\langle {\cal S} \rangle$ as a function of $V$. The sign shows a clear reduction around the known $V_{c}$ (denoted by the star). 

Figure~\ref{fig:3}$\,$\textbf{D} shows the spatial lattice size dependence of the sign, and Fig.~\ref{fig:3}$\,$\textbf{C}, once again,
a more `traditional' local variable, the derivative of the nearest-neighbor (NN) density-density correlation, $\langle n_{i}n_{j} \rangle_{\rm NN}$, with respect to $V$. In the CDW phase, increasing $V$ strengthens the staggered order, reducing the NN density correlations, and hence $- d \langle n_{i} n_{j} \rangle_{\rm NN} / d V$ is positive. Conversely, the effect is much smaller in the semimetal state, where the derivative is close to zero. The transition $V_{c}$ is characterized by a clear downturn in this quantity, which becomes progressively sharper as $L$ increases, as Figure~\ref{fig:3}$\,$\textbf{C} shows. This variable, thus, serves as a physical indicator of the QPT, allowing a comparison of Figs.~\ref{fig:3}\,$\textbf{C,D}$ to demonstrate the connection between the QCP and the behavior of $\langle {\cal S} \rangle$.  In this model, $\langle {\cal S}\rangle$ is well
enough behaved that a study of the finite-temperature CDW transition with DQMC is feasible~\cite{SM}, without having to resort to SP-free approaches~\cite{Hesselmann2016}.

%%%%%%%%%%%%%%%%%%%%%%%%%%%%%%%%%%%%%%%%%%%%%%%%%%%%%%%%%%%%%%%%%%
%%%%%%%%%%%%%%%%%%%%%%%%%%%%%%%%%%%%%%%%%%%%%%%%%%%%%%%%%%%%%%%%%%
\vskip0.10in 
{\bf \begin{center} IV. Square Lattice Hubbard Model \end{center}}  \label{sec:squarehub}
%%%%%%%%%%%%%%%%%%%%%%%%%%%%%%%%%%%%%%%%%%%%%%%%%%%%%%%%%%%%%%%%%%
%%%%%%%%%%%%%%%%%%%%%%%%%%%%%%%%%%%%%%%%%%%%%%%%%%%%%%%%%%%%%%%%%%

\begin{figure*}[htbp]
\includegraphics[width=1.8\columnwidth]{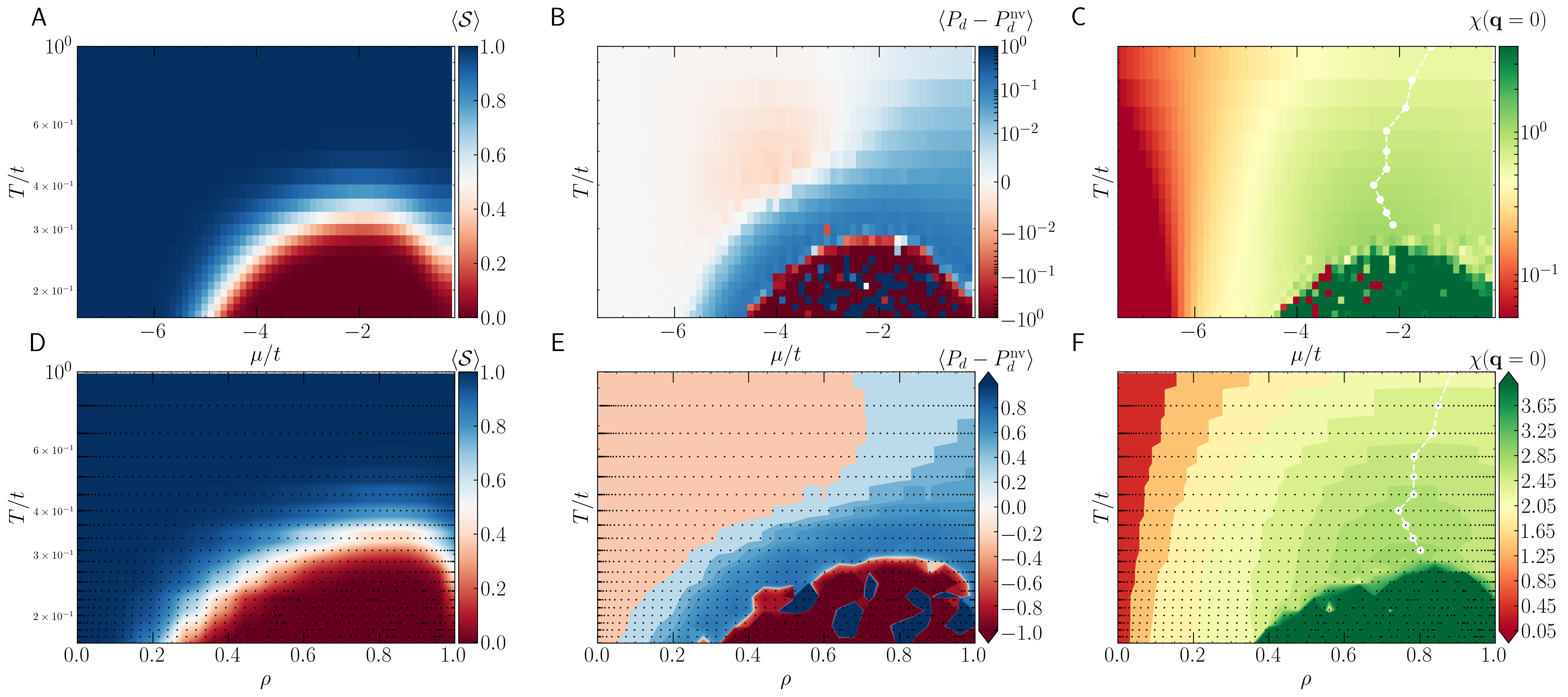}
% \vspace{-0.6cm}
\caption{\textbf{Square lattice Hubbard model.} (\textbf{A}) Temperature dependence of the average $\langle {\cal S}\rangle$ as function of the chemical potential $\mu/t$, for a lattice with $L=16$, $U/t = 6$ and next-near-neighbor hopping  $t^\prime/t = -0.2$, values chosen to be close to those in cuprate materials. (\textbf{B}) The $d$-wave pair susceptibility (subtracted from the non-vertex contribution) for the same parameters.  (\textbf{C}) The corresponding static spin susceptibility $\chi(\textbf{q}=0)$. The white markers describe its peak for values at which the average sign is large enough to allow a reliable calculation, which encompasses  the pseudogap regime. (\textbf{D}, \textbf{E} and \textbf{F}) display the corresponding diagrams when converting to the calculated average density. The black markers depict the actual average density extracted from the regular mesh of $\mu$ used in the upper panels, and where an interpolation of the data is performed. In all data, Trotter discretization is chosen as $\Delta\tau = 0.0625$.}
\label{fig:4}
\end{figure*}

The essential physics of the cuprate superconductors consists of antiferromagnetic order at and near one hole per CuO$_2$ cell, a superconducting dome upon doping, which typically extends to densities $0.6 \lesssim \rho \lesssim 0.9$ and a `pseudogap'/`strange metal' phase above the dome~\cite{Damascelli2003,Keimer2015}. There are many quantitative, experimentally-based, phase diagrams of different materials which determine the regions occupied by these phases~\cite{lee06}. Likewise, there are computational studies of individual $(\rho,T,U)$ points establishing magnetic/charge order~\cite{Jiang2019}, linear resistivity~\cite{Huang2019}, a reduction in the spectral weight for spin excitations~\cite{randeria92,tremblay06}, and $d$-wave pairing~\cite{Maier2005,Maier2016}. 

Here we reveal a `sign problem phase diagram' which bears significant resemblance to that of experiments. As is well known, the severity of the sign problem itself precludes determination of $d$-wave order in DQMC via `traditional' observables such as the associated correlation functions. However, Figure~\ref{fig:4}, based on the behavior of the sign itself, is suggestive. We report the average sign (Fig.~\ref{fig:4}$\,$\textbf{A}), the enhancement of the $d$-wave pairing susceptibility over its value in the absence of the pairing vertex~\cite{white89a} (Fig.~\ref{fig:4}$\,$\textbf{B}), and the uniform, static spin susceptibility $\chi(\textbf{q}=0)$ in both the $T/t-\mu/t$ (\textbf{A}--\textbf{C}) and $T/t-\rho$ (\textbf{D}--\textbf{F}) planes -- see SM~\cite{SM}.

The most salient features of this `sign phase diagram' are (i) the `dome' of vanishing $\langle {\cal S}\rangle$ which occurs in a range of densities $0.4 \lesssim \rho \lesssim 1$ as $T$ is lowered (Fig.~\ref{fig:4}$\,$\textbf{D}); (ii) the enhancement of $d$-wave pairing (Fig.~\ref{fig:4}$\,$\textbf{E}) surrounding the sign dome; and (iii), that
the magnetic properties are also linked to the $\langle {\cal S}\rangle$-dome: the trajectory tracing the
peak value of $\chi(q=0)$ as $T$ is decreased terminates precisely at the top of the dome
(Fig.~\ref{fig:4}$\,$\textbf{F}). 
In isolation, the comparisons of the behavior of the sign and the pairing and magnetic responses in the square lattice Hubbard model appear
likely to be merely coincidental. Indeed, the fact that the sign is worst precisely for optimal dopings has been previously remarked, but thought to be just `bad luck'~\cite{white89a,loh90,scalapino94,iglovikov15}.  However,
that the {\it known} QCP of the three models discussed in Secs.~I-III {\it can} be quantitatively linked to the behavior of $\langle {\cal S} \rangle$ suggest that the sign dome might actually be indicative
of the presence of $d$-wave superconductivity.

%%%%%%%%%%%%%%%%%%%%%%%%%%%%%%%%%%%%%%%%%%%%%%%%%%%%%%%%%%%%%%%%%%
%%%%%%%%%%%%%%%%%%%%%%%%%%%%%%%%%%%%%%%%%%%%%%%%%%%%%%%%%%%%%%%%%%
\vskip0.10in 
{\bf \begin{center} Discussion and Outlook  \end{center}} \label{sec:conclusions}
%%%%%%%%%%%%%%%%%%%%%%%%%%%%%%%%%%%%%%%%%%%%%%%%%%%%%%%%%%%%%%%%%%
%%%%%%%%%%%%%%%%%%%%%%%%%%%%%%%%%%%%%%%%%%%%%%%%%%%%%%%%%%%%%%%%%%

Early in the history of the study of the sign problem, a simple connection was noted between the fermionic physics and negative weights in auxiliary field QMC: If one artificially constructs two Hubbard-Stratonovich field configurations, one associated with two particle exchanging as they propagate in imaginary time and another with no exchange, one finds that the associated fermion determinants are negative in the former case, and positive in the latter~\cite{hirschmaybe}. This interesting observation, however, pertains to low density, that is, to the propagation of just two electrons. Another key observation is that the SP can be viewed as being proportional to the exponential of the difference of free energy densities of the original fermionic problem, and the one used with the weights in the Monte Carlo sampling taken to be positive, akin to a bosonic formulation of the problem~\cite{loh90,troyer05}. It highlights how intrinsic the SP is in QMC methods. A last important remark is that ordered phases are often associated with a reduction in the importance of configurations which scramble the sign. This is graphically illustrated in the snapshots of~\cite{hirsch82}. Although less crisp, similar effects are seen in auxiliary field QMC, for example in considering the evolution from the attractive Hubbard model to the Holstein model with decreasing phonon frequency $\omega_0$. Reducing $\omega_0$ acts to increase the effect of the phonon potential energy term $\hat P^2$ in $\hat H$, thereby straightening the auxiliary field in imaginary time. 

Here we have shown that the behavior of the average sign $\langle {\cal S} \rangle$ in DQMC simulations holds information concerning finite density thermodynamic phases and transitions between them: the quantum critical points in the semimetal to antiferromagnetic Mott insulator transition of Dirac fermions, 
%% the antiferromagnetic to singlet
%% transition in a multiband Hubbard Hamiltonian, and to 
the band insulator to metal to correlated insulator evolution of the ionic Hubbard Hamiltonian, and the QCP of spinless fermions (even though a sign-problem free formulation exists). Specifically, a rapid evolution of $\langle {\cal S} \rangle$ marks the positions of quantum critical points. We have chosen these models as representative examples of QCP physics of itinerant electrons which have been extensively studied in the condensed matter physics community, but believe the result to be general. In fact, in a model for frustrated spins in a ladder, using a completely different QMC method (stochastic series expansion), similar conclusions can be inferred~\cite{Wessel2017}, further corroborating this generality. 

Likewise, in the square lattice version of the $U(1)$ Hubbard model that we studied here, with an added $\pi$-flux, it can be shown that in the sign-problem free formulation, the QMC weights, when expressed in terms of the square of Pfaffians (Pf), holds similar information, namely, that $\langle {\rm sgn(Pf)}\rangle$ departs from 1 close to the QCP for this model~\cite{Goetz2021}. 
These results provide further evidence that the average sign of the QMC weights 
is inherently connected to the physics of the model in many unrelated models and methods, but an even broader study 
may 
be necessary to establish this conclusively.

Having established this connection in Hamiltonians with known physics, we have also presented a careful study of the sign problem for the Hubbard model on a 2D square lattice, which is of central interest to cuprate $d$-wave superconductivity. The intriguing `coincidence' that the sign problem is worst at a density $\rho \sim 0.87$, which corresponds to the highest values of the superconducting transition temperature, has previously been noted~\cite{white89a,loh90,scalapino94,iglovikov15}. It is worth emphasizing that we have not here presented any {\it solution} to the sign problem. However, our work does establish the surprising fact that $\langle {\cal S} \rangle$ can be used as an `observable' which can quite accurately locate quantum critical points in models like the spinful and spinless Hubbard Hamiltonians on a honeycomb lattice, and the ionic Hubbard model, and also provides a clearer connection between the evolution of the fermion sign and the strange metal/pseudogap and superconducting phases of the iconic
square lattice Hubbard model.

\noindent \paragraph{\bf Acknowledgments}

\noindent We acknowledge insightful discussions with S.-J. Hu and H.-Q. Lin. {\bf Funding:} R.T.S.~was supported by the grant DE‐SC0014671 funded by the U.S. Department of Energy, Office of Science. R.M.~acknowledges support from the National Natural Science Foundation of China (NSFC) Grants No. NSAF-U1930402, No.~11974039 and No.~12050410263. Computations were performed on the Tianhe-2JK at the Beijing Computational Science Research Center. {\bf Author contributions:} R.T.S proposed the original idea for the honeycomb lattice; R.M.~suggested its extension to the ionic and spinless fermion cases.  All authors considered the square lattice Hubbard model, performed numerical simulations, analyzed data, and co-wrote the manuscript. {\bf Competing interests:} Authors declare no competing interests. {\bf Data and materials availability:} All data needed to evaluate the conclusions in the paper are presented in the main text or the supplementary materials. Further raw data is available at~\cite{zenodo}.
 
\bibliography{inferringphysicssign}

\clearpage

\renewcommand{\theequation}{S\arabic{equation}}
\setcounter{equation}{0}

\onecolumngrid

\begin{center}

{\large \bf Supplementary Materials:
 \\ Quantum Critical Points and the Sign Problem }\\

\vspace{0.3cm}

\end{center}

\vspace{0.6cm}

% \twocolumngrid

%%%%%%%%%%%%%%%%%%%%%%%%%%%%%%%%%%%%%%%%%%%%%%%%%%%%%%%%%%%%%%%%
%%%%%%%%%%%%%%%%%%%%%%%%%%%%%%%%%%%%%%%%%%%%%%%%%%%%%%%%%%%%%%%%
\beginsupplement
%%%%%%%%%%%%%%%%%%%%%%%%%%%%%%%%%%%%%%%%%%%%%%%%%%%%%%%%%%%%%%%%
%%%%%%%%%%%%%%%%%%%%%%%%%%%%%%%%%%%%%%%%%%%%%%%%%%%%%%%%%%%%%%%%

In these Supplementary Materials we provide some additional context and history of the sign problem.  We also show the behavior of other observables across the transitions described in the main body of the paper, as well as exploring more carefully finite spatial size and Trotter effects in the evolution of $\langle {\cal S} \rangle$. 

\tableofcontents

%%%%%%%%%%%%%%%%%%%%%%%%%%%%%%%%%%%%%%%%%%%%%%%%%%%%%%%%%%%%%%%%
\vskip0.10in \noindent
\section{The Role of Quantum Simulations}
%%%%%%%%%%%%%%%%%%%%%%%%%%%%%%%%%%%%%%%%%%%%%%%%%%%%%%%%%%%%%%%%

The motivation for numerical solutions to the quantum many body problem is the intractability of analytic solutions except in limited circumstances. Indeed, the extension of the analytic solution of the single electron Hydrogen atom to two electrons is already problematic, as emphasized in the classic text of Bethe and Salpeter~\cite{bethe57}. As a consequence, quantum simulations approaches appeared in  chemistry~\cite{anderson75,ceperley81}, condensed matter~\cite{ceperley80,hirsch82}, nuclear~\cite{negele86,lynn19}, and high energy physics~\cite{kogut79,blankenbecler81,kogut83} almost as soon as computers became reasonably available for scientific work.
 
There is considerable methodological linkage in quantum simulation between these fields. For example, in the case of the DQMC method used in this work, the problem of treating the accumulation of round off errors at low temperatures was first developed in the nuclear physics community~\cite{sugiyama86} before being adapted to condensed matter. Likewise, the linear scaling methods of lattice gauge theory (LGT)~\cite{batrouni85,duane85,gottleib87} 
were soon adapted to condensed matter physics~\cite{scalettar86,scalettar87}, albeit with only
limited success owing to the extreme anisotropic (non-relativistic) nature of the condensed matter space-imaginary time lattices. Many of the algorithms, like Fourier Acceleration, which are crucial to LGT, were first implemented and tested for classical spin models~\cite{davies88}.
 
%%%%%%%%%%%%%%%%%%%%%%%%%%%%%%%%%%%%%%%%%%%%%%%%%%%%%%%%%%%%%%%%
\vskip0.10in \noindent
\section{Determinant Quantum Monte Carlo}
%%%%%%%%%%%%%%%%%%%%%%%%%%%%%%%%%%%%%%%%%%%%%%%%%%%%%%%%%%%%%%%%

The DQMC method is a specific type of AFQMC ~\cite{blankenbecler81,Buendia86,white89,chen92,assaad02,gubernatis16,alhassid17,hao19,he19}.
In DQMC, the partition function ${\cal Z}$ is expressed as a path integral
and the Trotter approximation is used to isolate the quartic terms,
\begin{align}
{\cal Z} = {\rm Tr} \, e^{-\beta \hat H}
= {\rm Tr} \, [ e^{-\Delta \tau \hat H} ]^{L_\tau}
\sim  {\rm Tr} \, [ e^{-\Delta \tau \hat H_t} e^{-\Delta \tau \hat H_U}  ]^{L_\tau}
\, ,
\end{align}
where $\hat H_t$ includes the hopping and chemical potential (together with all other bilinear terms in the fermionic operators), and $\hat H_U$ the on-site interactions, in the Hubbard Hamiltonian of Eq.~\eqref{eq:ham_spinful}. The latter are then decomposed via,
\begin{align}
e^{-\Delta \tau U (n_{i\uparrow} - \frac{1}{2}) 
(n_{i\downarrow} - \frac{1}{2})} 
= \frac{1}{2} e^{-U \Delta \tau/4} \sum_{s_i=\pm 1} e^{\lambda s_i (n_{i\uparrow} - n_{i\downarrow})}
\, ,
\label{eq:HS_spinful}
\end{align}
where ${\rm cosh} \, \lambda = e^{U \Delta \tau / 2}$. A Hubbard-Stratonovich variable (bosonic field) $s_i$ must be introduced at each spatial site $i$ and for each imaginary time slice
$U_{\Delta \tau}$. The prefactor $\frac{1}{2} e^{-U \Delta \tau/4} $ is an irrelevant constant which may be dropped.

The key observation is that the right hand side of  Eq.~\eqref{eq:HS_spinful} is now
quadratic in the fermions, so that the partition function ${\cal Z}$ is a trace over a product of quadratic forms of fermionic operators:
\begin{align}
{\cal Z} = \sum_{\{s_{i\tau} \}}{\rm Tr_{\uparrow}} \, [
e^{\vec c^{\, \dagger}_{\uparrow} K \vec c^{\phantom{\dagger}}_\uparrow}
e^{\vec c^{\, \dagger}_{\uparrow} P^1 \vec c^{\phantom{\dagger}}_\uparrow}
%% e^{\vec c^{\, \dagger}_{\uparrow} K \vec c^{\phantom{\dagger}}_\uparrow}
%% e^{\vec c^{\, \dagger}_{\uparrow} P^2 \vec c^{\phantom{\dagger}}_\uparrow}
\cdots
e^{\vec c^{\, \dagger}_{\uparrow} K \vec c^{\phantom{\dagger}}_\uparrow}
e^{\vec c^{\, \dagger}_{\uparrow} P^L_\tau \vec c^{\phantom{\dagger}}_\uparrow}
\, ]
\,\,\,  {\rm Tr_{\downarrow}} \, [
e^{\vec c^{\, \dagger}_{\downarrow} K \vec c^{\phantom{\dagger}}_\downarrow}
e^{- \vec c^{\, \dagger}_{\downarrow} P^1 \vec c^{\phantom{\dagger}}_\downarrow}
%% e^{\vec c^{\, \dagger}_{\downarrow} K \vec c^{\phantom{\dagger}}_\downarrow}
%% e^{- \vec c^{\, \dagger}_{\downarrow} P^2 \vec c^{\phantom{\dagger}}_\downarrow}
\cdots
e^{\vec c^{\, \dagger}_{\downarrow} K \vec c^{\phantom{\dagger}}_\downarrow}
e^{- \vec c^{\, \dagger}_{\downarrow} P^L_\tau \vec c^{\phantom{\dagger}}_\downarrow}
\, ]
\, ,
\label{eq:Z_tr}
\end{align}
with 
$ \vec c^{\, \dagger}_{\sigma} = 
 (\, c^{\, \dagger}_{\sigma 1  }
 \, c^{\, \dagger}_{\sigma 2  }
 \cdots
 \, c^{\, \dagger}_{\sigma N  }
 \, )^T$.
 %% and $ \vec c_{\sigma} =  \vec c^{\, \dagger}_{\sigma}   $
The matrix $K$ is the same for all time slices and contains $\mu \Delta \tau$ along its diagonal and $t\Delta \tau$ for sites connected by the hopping. (In the case of the ionic Hubbard model, the staggered site energy term also appears along the diagonal.) The matrices $P^\tau$ are diagonal, with $ P^\tau_{ii} = \lambda \sigma s_{i \tau}$ ($\sigma =\pm 1$ for $\uparrow$ and $\downarrow$). All matrices have dimension equal to the number of spatial lattice sites $N$.

The trace of a quadratic form such as Eq.~\eqref{eq:Z_tr} can be done analytically~\cite{blankenbecler81,white89,chen92,assaad02,gubernatis16,alhassid17,hao19,he19}, resulting in 
\begin{align}
{\cal Z} = \sum_{\{s_{i\tau} \}}
{\rm det} [ I + e^{K} e^{P^1}
%% e^{K} e^{P^2}
\cdots 
e^{K} e^{P^{L_\tau}} ]
 \,\, {\rm det} [ I + e^{K} e^{-P^1}
%% e^{K} e^{-P^2}
\cdots 
e^{K} e^{-P^{L_\tau}} ]
\, .
\label{eq:Z_det}
\end{align}
The expression for ${\cal Z}$ of Eq.~\eqref{eq:Z_det} contains no quantum operators, just the matrices $K, \{ P^\tau \}$ of the quadratic forms. Its calculation is thereby reduced to a classical Monte Carlo problem in which the sum over $\{ s_{i\tau} \}$ must be done stochastically with a weight equal to the product of the two determinants.  That these might become negative is the origin of the SP in DQMC.

Besides this spinful case, we have also investigated interacting spinless fermions. The problem formulation in QMC simulations is almost identical to that of the spinful case, but now the decoupling of the interactions $V$ on each nearest-neighbor bond $\langle i,j\rangle$ reads
\begin{align}
e^{-\Delta \tau V (n_i - \frac{1}{2}) (n_j - \frac{1}{2})} 
= \frac{1}{2} e^{-V \Delta \tau/4} \sum_{s_{ij}=\pm 1} e^{\lambda s_{ij} (n_i - n_j)}
\, ,
\label{eq:HS_spinless}
\end{align}
where ${\rm cosh} \, \lambda = e^{V \Delta \tau / 2}$. The Hubbard-Stratonovich variable $s$ now lives on the bonds, and its total number for the rhombus-shaped space-time lattices used here is $3L^2 L_\tau$. After a similar procedure for tracing out the fermions, one ends up with
\begin{align}
{\cal Z} = \sum_{\{s_{ij,\tau} \}}
{\rm det} [ I + e^{K} e^{P^1}
%% e^{K} e^{P^2}
\cdots 
e^{K} e^{P^{L_\tau}} ],
\label{eq:Z_det_spinless}
\end{align}
where the elements of the diagonal matrix $P^\tau$ are $P_{ii}^\tau = (-1)^i \lambda \sum_j s_{ij,\tau}$. A major difference from the spinful case is that the protection that the sign of the determinants of the matrices for up and down spin channels display in bipartite lattices by getting locked together~\cite{Hirsch1985,Iazzi2016} is no longer present. This can generically give rise to an even more detrimental SP, yet it allows us to systematically locate the QCP of Eq.~(\ref{eq:ham_spinless}) with a large accuracy.

%%%%%%%%%%%%%%%%%%%%%%%%%%%%%%%%%%%%%%%%%%%%%%%%%%%%%%%%%%%%%%%%
\vskip0.10in \noindent
\section{Physical Observables}
%%%%%%%%%%%%%%%%%%%%%%%%%%%%%%%%%%%%%%%%%%%%%%%%%%%%%%%%%%%%%%%%
The central object in the QMC simulations is the Green's function $\textbf{\rm G}^\sigma$ whose matrix elements are $G_{ij}^\sigma(\tau^\prime,\tau)=\langle c_{i\sigma}^{\phantom{\dagger}}(\tau^\prime) c_{j\sigma}^\dagger(\tau)\rangle$. By using Wick contractions and the fermionic anti-commutation relations one can define all quantities used in the main text. These are the double occupancy,
\begin{equation}
\langle n_{\uparrow\downarrow}\rangle = \langle n_\uparrow n_\downarrow\rangle ,
\end{equation}
the static susceptibility
\begin{equation}
\chi(\textbf{q}) = \frac{\beta}{N}\sum_{i,j}e^{{\rm i} \textbf{q}\cdot (\textbf{R}_i - \textbf{R}_j)} \langle (n_{i,\uparrow} - n_{i,\downarrow})(n_{j,\uparrow} - n_{j,\downarrow})\rangle,
\end{equation}
and the pair-susceptibility
\begin{equation}
P_\alpha = \int_0^\beta d\tau \langle \Delta_\alpha^{\phantom{\dagger}}(\tau)\Delta_\alpha^\dagger(0)\rangle,
\label{eq:pairsusc}
\end{equation}
with the momentum-dependent pair operator given by
\begin{equation}
\Delta_\alpha^\dagger = \frac{1}{N}\sum_{\textbf{k}} f_\alpha (\textbf{k})\,c_{\textbf{k}\uparrow}^\dagger c_{-\textbf{k}\downarrow}^\dagger.
\end{equation}
The form factors $f_\alpha (\textbf{k})$ describe the various symmetry channels investigated:
\begin{equation}
f_d (\textbf{k}) = \cos k_x - \cos k_y;\ \ f_{s^*} (\textbf{k}) = \cos k_x + \cos k_y; \ \ f_s(\textbf{k}) = 1,
\end{equation}
for $d$-wave, extended $s$-wave, and $s$-wave pairings, respectively.
In all data presented, we subtract the uncorrelated (non-vertex) contribution, $P_\alpha^{\rm nv}$, in which pairs of fermionic operators are first averaged before taking the product, i.e., terms in Eq.~(\ref{eq:pairsusc}) such as $\langle c_{i\downarrow}^\dagger(\tau)c_{j\downarrow}^{\phantom{\dagger}}(0)c_{l\downarrow}^\dagger(\tau)c_{m\downarrow}^{\phantom{\dagger}}(0)\rangle$ get replaced by their decoupled contributions $\langle c_{i\downarrow}^\dagger(\tau)c_{j\downarrow}^{\phantom{\dagger}}(0)\rangle \, \langle c_{l\downarrow}^\dagger(\tau)c_{m\downarrow}^{\phantom{\dagger}}(0)\rangle$. 

Other quantities used exclusively in this supplemental material are introduced in the corresponding sections.
%%%%%%%%%%%%%%%%%%%%%%%%%%%%%%%%%%%%%%%%%%%%%%%%%%%%%%%%%%%%%%%%
\vskip0.10in \noindent
\section{Sign Problem: General Importance}
%%%%%%%%%%%%%%%%%%%%%%%%%%%%%%%%%%%%%%%%%%%%%%%%%%%%%%%%%%%%%%%%

In lattice gauge theory, the SP is triggered by a non-zero chemical potential $\mu$.
Attempts to solve, or reduce, the SP include analytic continuation from complex to real chemical potentials $\mu$~\cite{alford99,toulouse16}, Taylor expansion about zero $\mu$~\cite{deforcrand02,gavai03}, re-weighting approaches~\cite{allton02}, complex Langevin methods~\cite{adami01,berger19}, and Lefshetz thimble techniques~\cite{witten10,cristoforetti12}. For a review of the SP in LGT, see Ref.~\cite{banuls19}. A similar litany of papers, ideas, and new methods characterizing, ameliorating, or solving the SP can be found in the condensed matter~\cite{loh90,ortiz93,zhang97,chandrasekharan99,henelius00,bergkvist03,troyer05,nyfeler08,nomura14,mukherjee14,shinaoka15,kaul15,iglovikov15,he19,fukuma19,ulybyshev20,kim20}, nuclear physics ~\cite{wiringa00,alhassid17,lynn19,gandolfi19}, and quantum chemistry~\cite{umrigar07,shi14,roggero18,rothstein19} communities.
These approaches differ in detail, depending on whether they are addressing the SP in real space versus lattice models, quantum spin versus itinerant fermion, etc.

One of the dominant themes in atomic and molecular (AMO) physics over the last decade has been the possibility that ultracold atoms in an optical lattice might serve as emulators of fundamental models of condensed matter physics~\cite{bloch05,esslinger10,bloch12,tarruell18,schafer20}. The initial motivation was the possibility that simplified model Hamiltonians could be more precisely realized in the AMO context than in the solid state, where materials `complications' are unavoidable and often significant, if not dominant. However, it was quickly understood that an equally important advantage was that, because of the SP, the solution of these simplified models was possible only at temperatures which were well above those needed to access phenomena like $d$-wave superconductivity.
Thus it is fair to say that the SP has been a significant driver of the enormous efforts and progress in this domain of AMO physics.

A similar theme is present in quantum computing~\cite{ortiz01,brown10,preskill18,clemente20}, which promises the general possibility of solving problems much more rapidly (`quantum supremacy') than with a classical computer. The exponential scaling time of solutions of model Hamiltonians in the presence of the SP offers one of the most significant targets for such endeavors~\cite{Steudtner2018,Arute2020}.

%%%%%%%%%%%%%%%%%%%%%%%%%%%%%%%%%%%%%%%%%%%%%%%%%%%%%%%%%%%%%%%%
\vskip0.10in \noindent
\section{The sign problem in other QMC algorithms}
%%%%%%%%%%%%%%%%%%%%%%%%%%%%%%%%%%%%%%%%%%%%%%%%%%%%%%%%%%%%%%%%

In exploring the linkage between the SP and critical properties, we have focused exclusively on DQMC.  As noted in the previous discussion, the SP occurs also in a plethora of QMC methods:  world-line quantum Monte Carlo (WLQMC)~\cite{kawashima04}, Greens Function (Diffusion) Monte Carlo (GFMC)~\cite{reynolds90}, in dynamical mean field theory (DMFT) and its cluster extensions (including continuous time approaches)~\cite{georges96,beard96,hettler00,kotliar01,rubtsov05,maier05,kyung06,gull08,gull11}, and in diagrammatic QMC~\cite{prokofev96,vanhoucke10,rossi17,rohringer18}. It would be interesting to explore these situations as well, checking the direct correlation between the SP and regimes of quantum critical behavior. Regarding WLQMC, it is known that the onset of the sign problem is at much higher temperature than in DQMC, and occurs due to particle exchange in the world-lines.  Indeed, this provides one of the earliest examples of the dependence of the sign problem on the algorithm used.  It seems probable that the restriction of WLQMC to only very high temperatures ($T/W \gtrsim 1/5$ is quite typical) might preclude the possibility of an association of the sign problem with interesting low temperature correlations and transitions. It is worth noticing that an attempt to reconcile WLQMC and DQMC was put forward by introducing more generic Hubbard-Stratonovich transformations~\cite{Hirsch1986}; it has been further argued that the SP in WLQMC has two origins: one from the fact that Slater determinants are anti-symmetric sums of world-line configurations and another intrinsic, akin to the one in DQMC, that has been claimed having topological origin~\cite{Iazzi2016}. In the same way, alternative choices of Hubbard-Stratonovich transformations in DQMC~\cite{batrouni90,chen92} significantly worsen the sign problem. DMFT~\cite{georges92,jarrell92,georges96}, on the other hand, is at the opposite end of the spectrum, with a sign problem which is greatly reduced relative to that of DQMC.  Unfortunately, its cluster extensions, using QMC as the cluster solver, exhibit an increasingly serious SP at low enough temperatures as the number of points in the momentum grid increases~\cite{jarrell01,maier05,kyung06}.

\begin{figure}[b]
\includegraphics[scale=0.33]{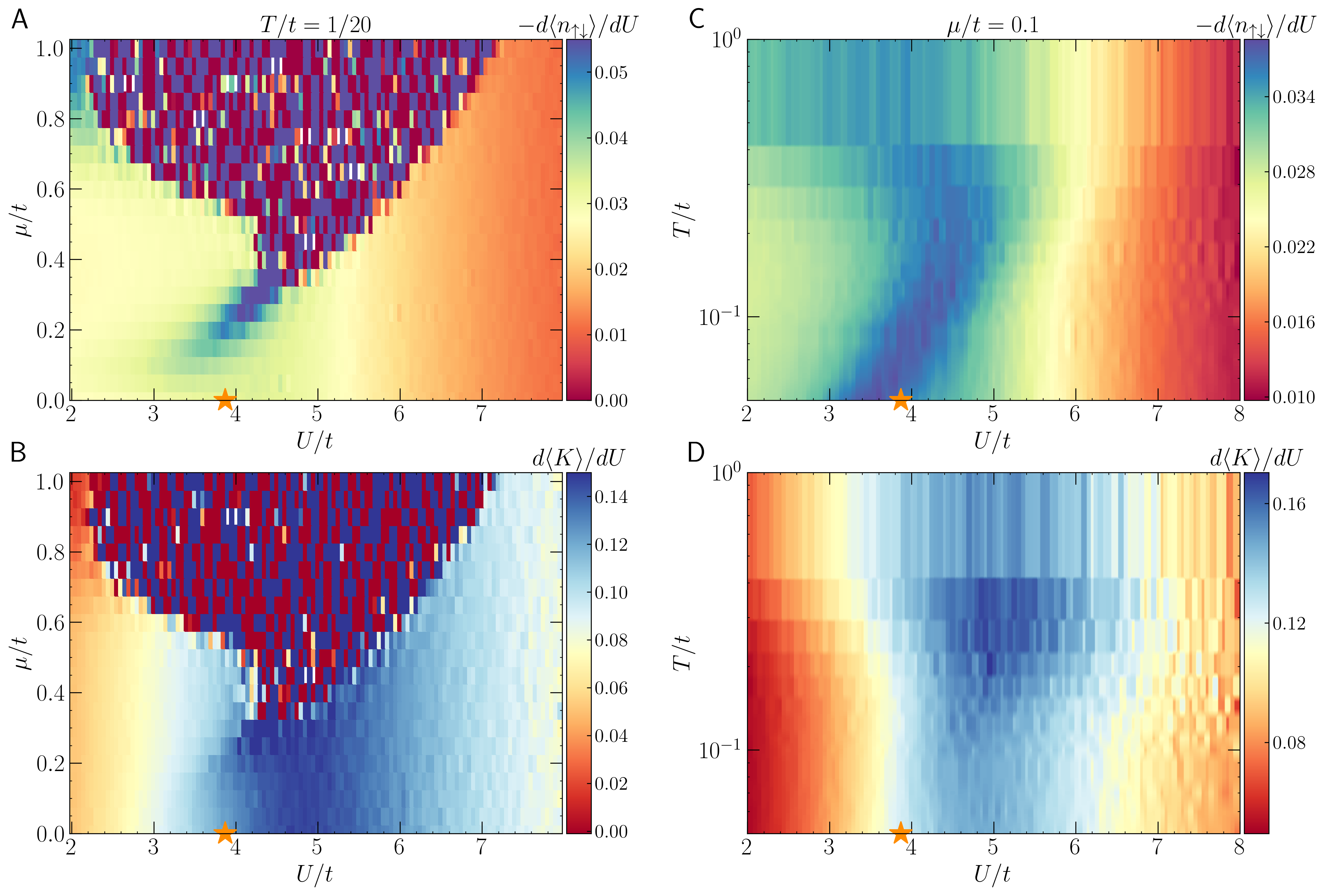}
\caption{\textbf{Local observables in the vicinity of the quantum critical point on the SU(2) honeycomb Hubbard model.} The derivative of the double occupancy (\textbf{A}) and the kinetic energy  (\textbf{B}) at $T/t = 1/20$ when approaching the quantum critical point at $\mu\to0$ and $U_c\simeq3.85t$. As in the main text for other models, the noisy data at large chemical potentials denotes the regime of small $\langle{\cal S}\rangle$ with a vanishing signal-to-noise ratio. (\textbf{C})  and (\textbf{D}) display the same quantities when approaching $T/t\to0$, with a small chemical potential $\mu/t=0.1$. As in the $T=0$ results of Ref.~\cite{meng10}, the derivative of the kinetic energy displays a peak within the AFMI phase. Our data show this persists to finite temperatures. The lattice size used is $L=9$, and all data are averaged among 20 independent runs.}
\label{fig:spinful_SM}
\end{figure}

We finish by noting that while we have argued that when a QCP is present, the SP provides quantitative information about its location, the converse is {\it not} necessarily true: a SP can exist even in the absence of a QCP. In particular, in WLQMC, free fermions have a SP in $D>1$ without possessing any sort of phase transition. As noted above, DQMC is SP free when the interactions vanish, so this simple counter-example is not present in that algorithm.

%%%%%%%%%%%%%%%%%%%%%%%%%%%%%%%%%%%%%%%%%%%%%%%%%%%%%%%%%%%%%%%%
\vskip0.10in \noindent
\section{The sign problem in other Hamiltonians}
%%%%%%%%%%%%%%%%%%%%%%%%%%%%%%%%%%%%%%%%%%%%%%%%%%%%%%%%%%%%%%%%
 
In our work, the spinless fermion Hamiltonian on a honeycomb lattice offered a particularly concrete case where a sign problem free approach allows a detailed study of quantum critical behavior.  We exploited this as a way to make a very quantitative test of the connection of the SP to the location of the QCP. In addition to exploring other algorithms, a promising further line of inquiry is to turn on a  chemical potential in other `sign problem free' models~\cite{chandrasekharan10,berg12,wang15,li16,li19}. The Kane-Mele-Hubbard model would be especially interesting since it presents a framework to understand the transition from topological phases (quantum spin Hall insulator) towards a (topologically trivial) Mott insulator with antiferromagnetic order~\cite{Hohenadler2012,Bercx2014,Toldin2015}. The Hubbard-Stratonovich transformation is slightly more complicated, and in fact a `phase problem' appears instead. In exploring this competition between ordered and topological phases, the study of the Haldane-Hubbard model presents as a challenging case, since due to the absence of time-reversal symmetry, it gives rise to a severe SP in the simulations~\cite{Imriska2016}. Whether similar analysis as conducted here can help in locating the QCP associated to a topological transition in system sizes exceeding the ones amenable to ED~\cite{Vanhala2016,Shao2021} is yet an open question left for future studies.

A further possibility for future work includes situations where disorder drives a quantum phase transition~\cite{huscroft97}. This is of particular interest because different types of disorder can either possess particle-hole symmetry or not~\cite{denteneer01}, and this dichotomy is known to be linked both to the presence of  the sign problem as well as to the occurrence of metal to insulator transitions~\cite{denteneer99,denteneer03}. Thus disordered systems might provide an especially rich arena to explore the connection between the SP and the underlying physics of the Hamiltonian.

%%%%%%%%%%%%%%%%%%%%%%%%%%%%%%%%%%%%%%%%%%%%%%%%%%%%%%%%%%%%%%%%
\vskip0.10in \noindent
\section{More details on the Spinful Hubbard Model on a Honeycomb Lattice (main text, Sec.~I)}
%%%%%%%%%%%%%%%%%%%%%%%%%%%%%%%%%%%%%%%%%%%%%%%%%%%%%%%%%%%%%%%%
\paragraph{Energy and Double Occupation.---}The quantum critical point separating the paramagnetic semimetal and the antiferromagnetic insulator (AFMI) phases of the honeycomb Hubbard hamiltonian has been characterized using observables ranging from the magnetic structure factor to the conductivity. In the main text, we focused on using the average sign as a signal of the QCP. Here we provide context by showing some of the  traditional measurements. More detailed results are contained in the literature~\cite{sorella92,paiva05,herbut06,meng10,giuliani10,ma11,clark11,sorella12,raczkowski20}.

Figure~\ref{fig:spinful_SM} shows the derivatives of the average kinetic energy $\langle K \rangle = \frac{-t}{2L^2}\sum_{\langle i,j\rangle,\sigma} (c_{i \sigma}^{\dagger} c_{j \sigma}^{\phantom{}} + c_{j\sigma}^{\dagger} c_{i \sigma}^{\phantom{}})$ and double occupation $\langle n_{\uparrow \downarrow} \rangle$ with respect to $U$ across the honeycomb lattice transition from semimetal to AFMI.  Both show clear signals in the vicinity of $U_c$.  The accurate indication of antiferromagnetic long range order requires a careful finite size scaling analysis of the antiferromagnetic structure factor, which can be found in Refs.~\cite{paiva05,meng10,sorella12}.

\begin{figure}[b]
\includegraphics[width=0.6\columnwidth]{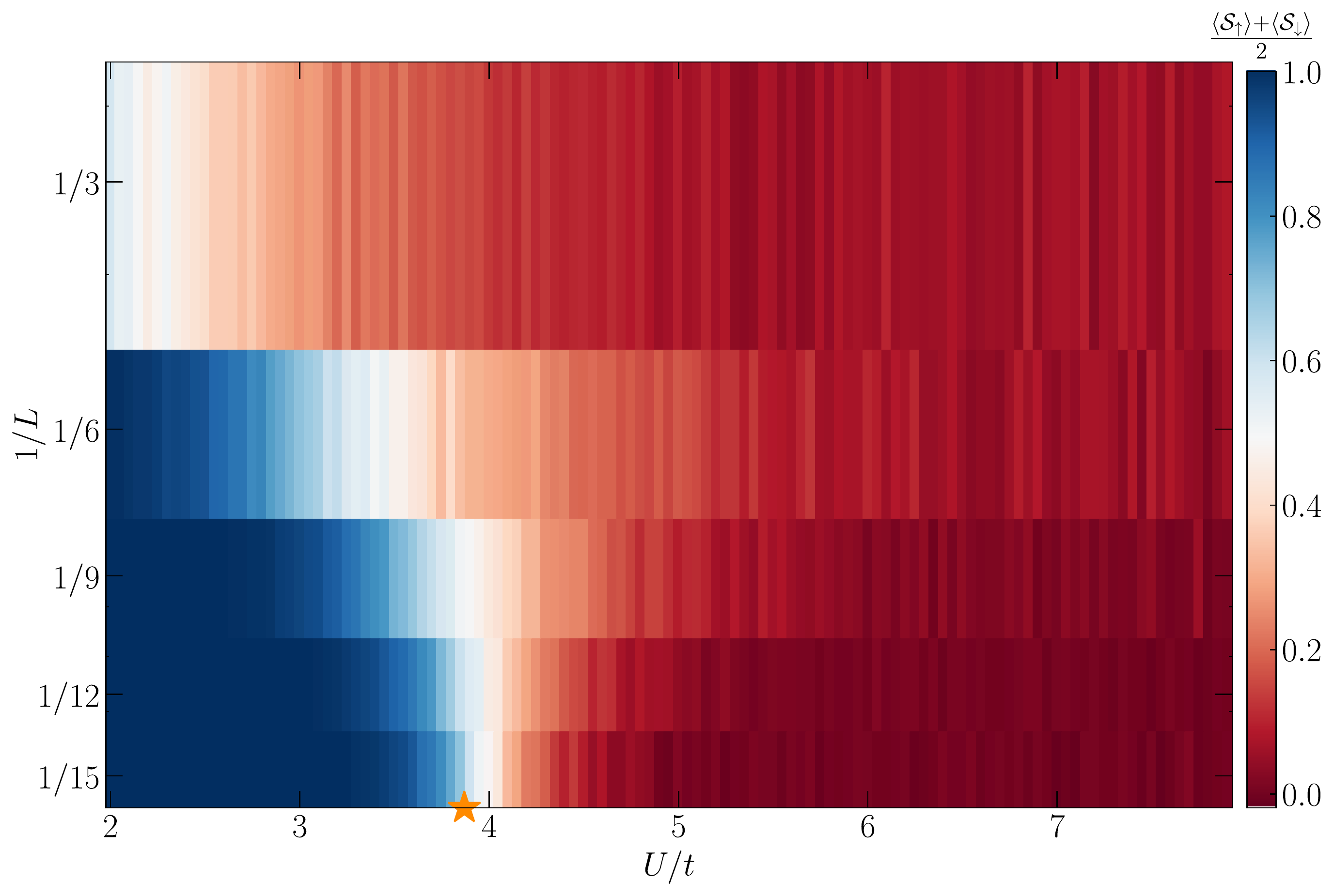}
\caption{\textbf{Spin resolved sign for the SU(2) Hubbard model on the honeycomb lattice.} The  scaling with the inverse of the linear system size $L$ of the average sign of individual  determinants in the case there is no sign problem: $\mu/t = 0$. The star marker depicts the best known estimation of the critical value $U_c/t = 3.869$~\cite{sorella12} at the ground state for the onset of the Mott insulating phase with AFM order. All data are extracted at $T/t = 1/20$, with $\Delta\tau = 0.1$.}
\label{fig:spin_resolv_sign}
\end{figure}

\paragraph{Individual spin channel average sign in the Spinful Hubbard Model on a Honeycomb Lattice.---} In the main text, we have demonstrated how the average sign can be used as a `tracker' of quantum critical behavior. In the case of models within regimes where a sign problem is absent, e.g., for an SU(2) Hubbard model on a bipartite lattice, this ability is no longer available if the chemical potential $\mu = 0$, since the determinants of the two spin species always have the same sign so that their product is positive.  This can be proven by considering a staggered particle-hole transformation (PHT) $c^{\dagger}_{i\downarrow} \rightarrow (-1)^i c^{\phantom{\dagger}}_{i\downarrow}$ on the down spin species.  Here $(-1)^i = +1(-1)$ on sublattice A(B). Under the PHT, the kinetic energy matrix $K$ of Eqs.~\eqref{eq:Z_tr},\eqref{eq:Z_det} remains invariant, but the  matrices $P^\tau$ in the down spin trace change sign, making the down spin determinant the same as the  up spin determinant, up to a positive factor $e^{\lambda \sum_{i \tau} s_{i\tau}}$~\cite{Hirsch1985}.
 
While the {\it product} of the determinants is always positive in this situation, the QCP remains imprinted in the average sign of the determinants for {\it individual} spin components $\langle {\cal S}_\sigma\rangle$. To illustrate this, we consider the first model used in the main text, the repulsive spinful Hubbard model on the (bipartite) honeycomb lattice. Figure~\ref{fig:spin_resolv_sign} plots $\langle {\cal S}_\sigma\rangle$  ($\sigma=\uparrow,\downarrow$) at fixed $T/t=1/20$.  The individual signs are largely positive in the metallic phase, but rather abruptly change to equally positive and negative ($\langle {\cal S} \rangle \sim 0$) in the AFMI phase $U>U_c$. The match of the transition in the sign and the position of the QCP becomes increasingly precise in the thermodynamic limit $1/L \rightarrow 0$. The sharpness of the drop in $\langle {\cal S}_\sigma\rangle$ with increasing system sizes is suggestive of a possible scaling form for this quantity. Preliminary data, to be presented elsewhere, indicates a scaling with critical exponents compatible with the ones obtained from physical observables~\cite{Assaad2013}.

%%%%%%%%%%%%%%%%%%%%%%%%%%%%%%%%%%%%%%%%%%%%%%%%%%%%%%%%%%%%%%%%
\vskip0.10in \noindent
\section{More details on the Ionic Hubbard Hamiltonian (main text, Sec.~II)}
%%%%%%%%%%%%%%%%%%%%%%%%%%%%%%%%%%%%%%%%%%%%%%%%%%%%%%%%%%%%%%%%

\paragraph{Finite-size effects.---} The ionic Hubbard model presents a unique situation in our study: instead of displaying a quantum critical point at half-filling, it exhibits a quantum critical regime, associated with a correlated metal (CM) phase~\cite{paris07,bouadim07,fabrizio99,kampf03,manmana04,garg06,paris07,bouadim07,craco08,garg14,bag15}.  We argued in the main text that this phase, sandwiched between the band-insulator (BI) at large $\Delta$, and the Mott insulator (MI) at large $U$, can be indicated by a vanishing average sign in the DQMC simulations. We now explore the influence of finite-size effects on those phase boundaries, for a specific value of the staggered potential $\Delta/t=0.5$. 

\begin{figure*}[th!]
\includegraphics[width=1\columnwidth]{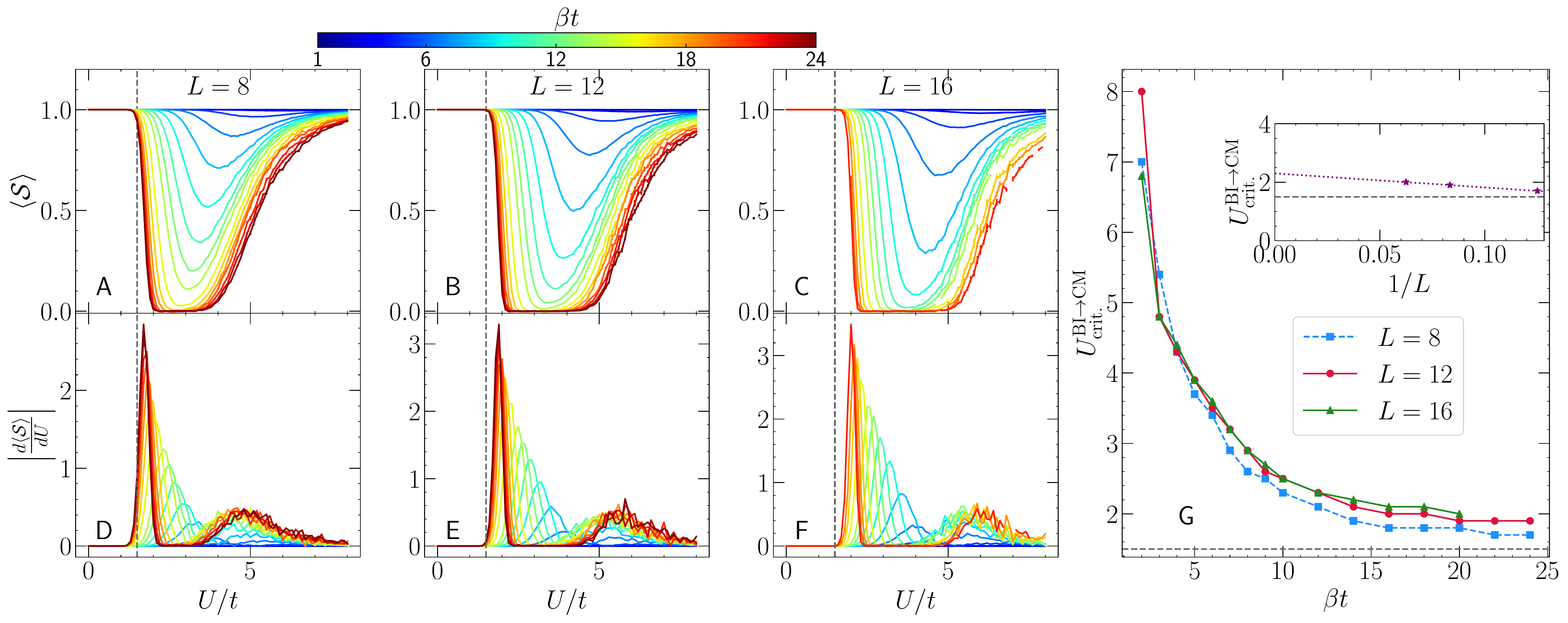}
\caption{\textbf{Finite-size analysis: Ionic Hubbard model.} (\textbf{A, B, C}) The average sign $\langle{\cal S}\rangle$ for decreasing temperatures (growing $\beta = 1/T$) as a function of the Hubbard interaction $U/t$ in a lattice with $L=8, 12$ and 16 with a staggered potential $\Delta/t=0.5$. (\textbf{D,E,F}) The corresponding magnitude of the derivative of $\langle{\cal S}\rangle$ in respect to $U$ at each lattice size. (\textbf{G}) Peak position of the derivative of the average sign with different inverse temperatures $\beta$; the inset presents an extrapolation to the thermodynamic limit of the estimated transition from the BI to the CM phase, as inferred by the first peak of $|d\langle {\cal S}\rangle/dU|$. The gray dashed lines in all panels display the corresponding transition value obtained in Ref.~\cite{bouadim07}. All data are averaged over 24 independent runs, with $\Delta\tau = 0.1$.}
\label{fig:ionic_SM}
\end{figure*}

Figure~\ref{fig:ionic_SM} indicates that the first transition which occurs upon increasing  $U/t$ from zero, that from BI to CM, is well marked by a fast drop of $\langle{\cal S}\rangle$ at lower temperatures. A quantitative estimation of the transition point can be extracted by differentiating the average sign with respect to the interaction strength, $d \langle {\cal S} \rangle / d U$ (lower panels in Fig.~\ref{fig:ionic_SM}).  As the temperature is lowered, the peak position quickly approaches the best known values of the transition for this set of parameters~\cite{bouadim07}, see Fig.~\ref{fig:ionic_SM}(\textbf{G}). The system size dependence is reasonably small. The second transition, from CM to AFI, on the other hand, displays characteristics reminiscent of a crossover for the system sizes and temperatures investigated. The estimate given by the peak of the average sign also displays a stronger dependence on $L$, and, overall, is larger than the value of the position of the metal-AF transition at this $\Delta$ obtained in Ref.~\cite{bouadim07}. It is worth mentioning that these values in the existing literature were extracted at smaller lattice sizes than the largest $L$ used here.  A finite size extrapolation of the `traditional' correlations used to obtain $U_c$ for the metal-AF transition, similar to the one we perform here, would be useful to undertake.

\begin{figure*}[th!]
\includegraphics[width=0.6\columnwidth]{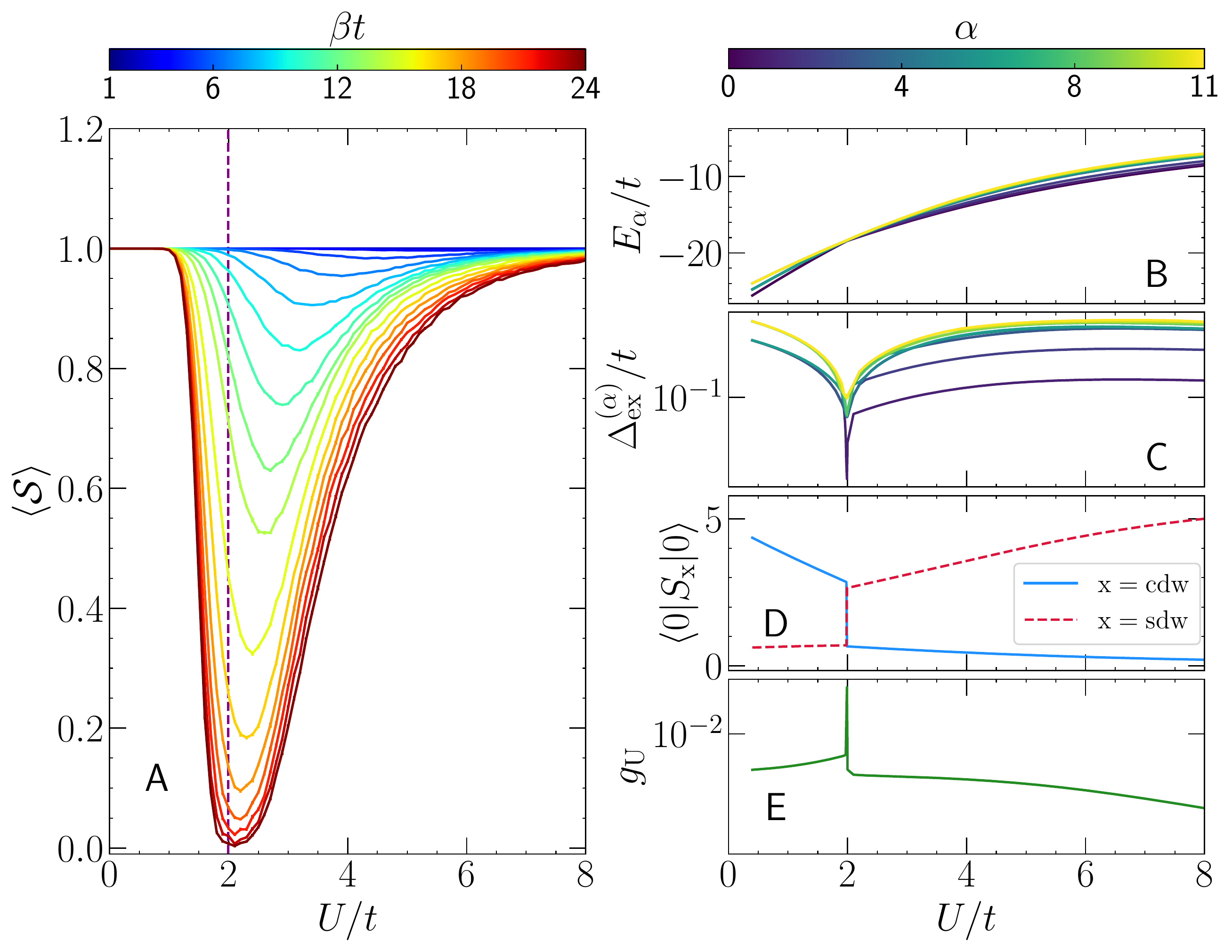}
\caption{\textbf{QMC vs. ED on a 4x4 lattice.} (\textbf{A}) The average sign extracted from the DQMC calculations $\langle{\cal S}\rangle$ for increasing inverse temperatures $\beta = 1/T$ as a function of the Hubbard interaction $U/t$ in a lattice with $L=4$ and a staggered potential $\Delta/t=0.5$. (\textbf{B--E}) Data extracted with the ED, including: (\textbf{B}) The low-lying eigenspectrum $E_\alpha$ at the zero-momentum sector $\mathbf{k} = (0,0)$; (\textbf{C}) The excitation gaps $\Delta_{\rm ex}^{(\alpha)}$; (\textbf{D}) the spin and charge structure factors and (\textbf{E}) the fidelity metric under variations of the interaction magnitudes $U$, using $dU=10^{-3}t$. In (\textbf{A}), data are averaged over 24 independent runs, with $\Delta\tau = 0.1$.}
\label{fig:ionic_QMC_ED_comparison_SM}
\end{figure*}

\paragraph{QMC vs. ED.---} A valuable test of the conjecture that $\langle {\cal S}\rangle$ tracks a quantum phase transition (or regime) can be made by comparing QMC results with exact ones, obtained at smaller lattice sizes. For this purpose, we contrast in Fig.~\ref{fig:ionic_QMC_ED_comparison_SM} the average sign in a lattice with $L=4$ with numerical results obtained from exact diagonalization (ED). At this small lattice size, the quantum critical region shrinks, and at the lowest temperatures studied ($T/t = 1/24$) $\langle{\cal S}\rangle$ displays a sharp dip at around $V/t\simeq 2$. Turning to the ED results, we probe the transition via the analysis of the low lying spectrum $E_\alpha$ ($E_0$ is the ground-state), the many-body excitation gaps $\Delta_{\rm ex}^{(\alpha)} = E_\alpha - E_0$, the spin and charge staggered structure factors, $S_{\rm sdw} = (1/N)\sum_{i,j} (-1)^\eta \langle (n_{i\uparrow} - n_{i\downarrow})(n_{j\uparrow} - n_{j\downarrow})\rangle$ and $S_{\rm cdw} = (1/N)\sum_{i,j} (-1)^\eta \langle (n_{i\uparrow} + n_{i\downarrow})(n_{j\uparrow} + n_{j\downarrow})\rangle$ [$\eta = +1\  (-1)$ when $i,j$ belong to the same (different) sublattices], and, the fidelity metric $g_{\tiny U} = \frac{2}{N}\frac{1-|\langle \Psi_0(U)|\Psi_0(U+dU)\rangle|}{dU^2}$. This last quantity displays a peak whenever one crosses a quantum phase transition for the parameters $U$ and $U+dU$. These results describe a single transition in the range of parameters investigated, displaying a first-order character, given the level crossings shown in Fig.~\ref{fig:ionic_QMC_ED_comparison_SM}(\textbf{B}) or a vanishing excitation gap $\Delta_{\rm ex}^{(1)} = E_1 - E_0$ at $U/t \simeq 1.99$ [Fig.~\ref{fig:ionic_QMC_ED_comparison_SM}(\textbf{C})]. In turn, the fidelity metric displays a sharp peak at this interaction value [Fig.~\ref{fig:ionic_QMC_ED_comparison_SM}(\textbf{E})], and the structure factors computed at the ground-state swap its characteristics, from a charge- to a spin-ordered one [Fig.~\ref{fig:ionic_QMC_ED_comparison_SM}(\textbf{D})].
It is an open question of whether one is able to capture the intermediate correlated metal phase in exact methods such as ED.

Stepping back from the technical details, the central message of 
Fig.~\ref{fig:ionic_QMC_ED_comparison_SM} is extending the evidence presented in the main text 
that the sign problem metric for the QCP of panel \textbf{A}
lines up well with those of the `traditional observables' in panels \textbf{B-E}.

%%%%%%%%%%%%%%%%%%%%%%%%%%%%%%%%%%%%%%%%%%%%%%%%%%%%%%%%%%%%%%%%
\vskip0.10in \noindent
\section{More details on Spinless Honeycomb Hubbard (main text, Sec.~III)}
%%%%%%%%%%%%%%%%%%%%%%%%%%%%%%%%%%%%%%%%%%%%%%%%%%%%%%%%%%%%%%%%

\paragraph{The finite-temperature transition.---} As we have argued in the main text, the interacting spinless fermion Hamiltonian has a special property in AFQMC simulations: with an appropriate choice of the basis one uses to write the fermionic matrix, it has been proven that the sign problem can be eliminated~\cite{ZiXiang2015,li16,li19}. Nonetheless, using a standard single-particle basis, where the sign problem is manifest, we demonstrated in Fig.~\ref{fig:3} that $\langle{\cal S}\rangle$ can be used as a way to track the quantum phase transition. Concomitantly, we have shown that a local observable (the derivative of the nearest neighbor density correlations with respect to the interactions $V$) exhibits a steep downturn once the quantum (i.e.~zero temperature) phase transition is approached.

\begin{figure}[th!]
\includegraphics[width=0.7\columnwidth]{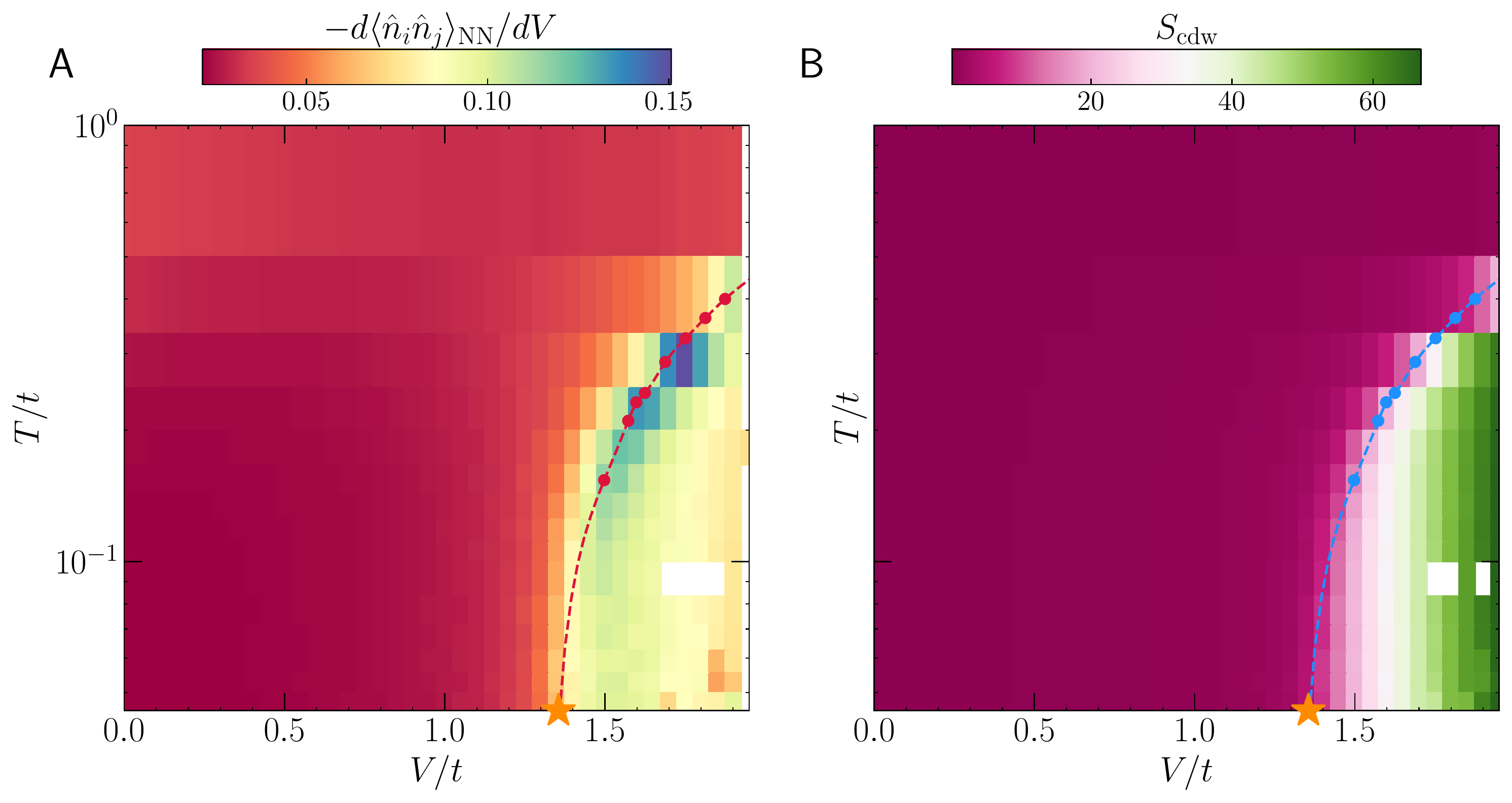}
\caption{\textbf{Finite-temperature transition for the spinless fermion on the honeycomb lattice.} As a by-product of the analysis of the temperature dependence for average sign, we notice that other than in regimes very close to the QCP, $\langle {\cal S}\rangle$ is very close to 1. (\textbf{A}) The derivative of the nearest-neighbor density correlations in the $T/t$ vs. $V/t$ plane and (\textbf{B}) the CDW structure factor for a lattice with $L=18$. In both plots, the markers are the continuous-time QMC results extracted after finite-size scaling in Ref.~\cite{Hesselmann2016}. Imaginary-time discretization is fixed at $\Delta\tau = 0.1$, and all data is obtained as an average of 20 independent runs.}
\label{fig:spinless_hc_SM}
\end{figure}

As a by-product of this analysis, we use our original approach based on the standard BSS algorithm in the standard fermionic basis to show that one can also obtain an estimation of the \textit{finite-temperature} transition (pertaining to universality class of the 2D Ising model) with a relatively large accuracy, if the system is not too close to the quantum critical point, see Fig.~\ref{fig:spinless_hc_SM}. We compute both the derivative of the nearest-neighbor density correlations as well as the CDW structure factor, i.e., a summation of all density-density correlations with a $+1$ $(-1)$ for sites belonging to the same (different) sublattice, on the largest lattice size we have investigated, $L=18$. These finite-size results for $T_c$ are in good agreement with recent results obtained after system size scaling of data extracted with continuous-time QMC methods~\cite{Wang2014,Hesselmann2016}, where the sign problem is absent.

\begin{figure}[th!]
\includegraphics[width=0.6\columnwidth]{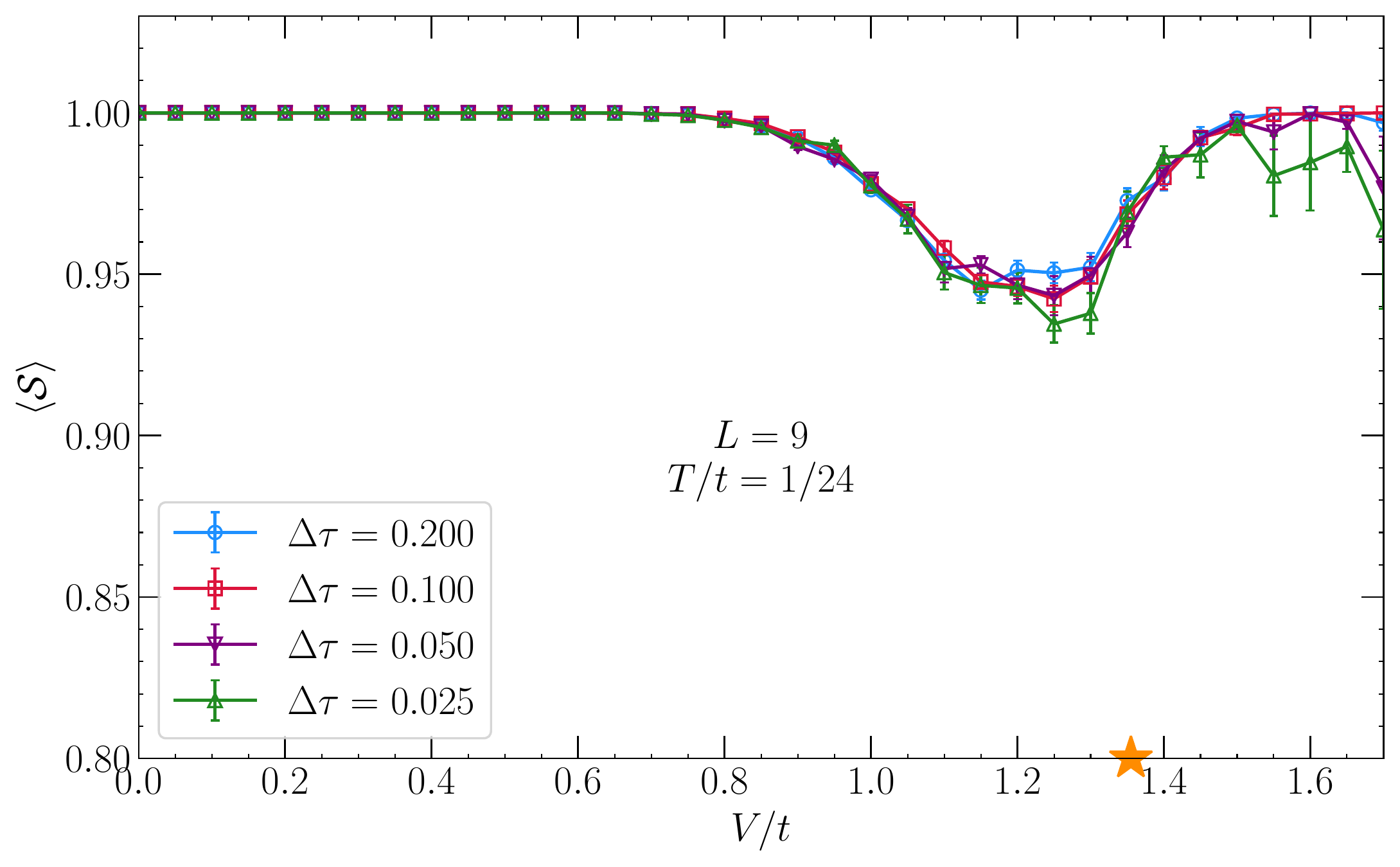}
\caption{\textbf{The U(1) model on the honeycomb lattice: Average sign dependence on the imaginary time-discretization.} The dependence of $\langle{\cal S}\rangle$  on the imaginary-time discretization $\Delta\tau$ with growing interactions $V/t$, in a lattice with $L = 9$. Data is extracted as an average of 24 independent runs, for a temperature $T/t=1/24$. The best known value for the Mott insulating transition~\cite{ZiXiang2015} ($V_c/t=1.355$) is signalled by the star marker.}
\label{fig:spinless_dtau_sign}
\end{figure}

An important outcome of these results is that they imply that the mark that a phase transition leaves on the average sign is restricted to quantum phase transitions, rather than thermal ones, as we show above.

\paragraph{Imaginary-time discretization.---} 

Here we consider the effect of the imaginary-time discretization $\Delta\tau$ on the average sign, for a fixed inverse temperature and show that `Trotter errors'~\cite{fye86} do not affect our conclusions. Figure~\ref{fig:spinless_dtau_sign} shows the average sign for a fixed lattice size $L=9$ and $\Delta\tau$ ranging from 0.025 to 0.2 with a fixed temperature $T/t=1/24$. The drop in $\langle{\cal S}\rangle$ when approaching $V_c$ is indicative of the QCP, but using a more dense imaginary-time discretization does not render substantial changes in the average sign; similar behavior was observed in other models studied.

%%%%%%%%%%%%%%%%%%%%%%%%%%%%%%%%%%%%%%%%%%%%%%%%%%%%%%%%%%%%%%%%
\vskip0.10in \noindent
\section{More details for the homogeneous Hubbard model on the square lattice (main text, Sec.~IV)}
%%%%%%%%%%%%%%%%%%%%%%%%%%%%%%%%%%%%%%%%%%%%%%%%%%%%%%%%%%%%%%%%

\begin{figure*}[th!]
\includegraphics[width=1.\columnwidth]{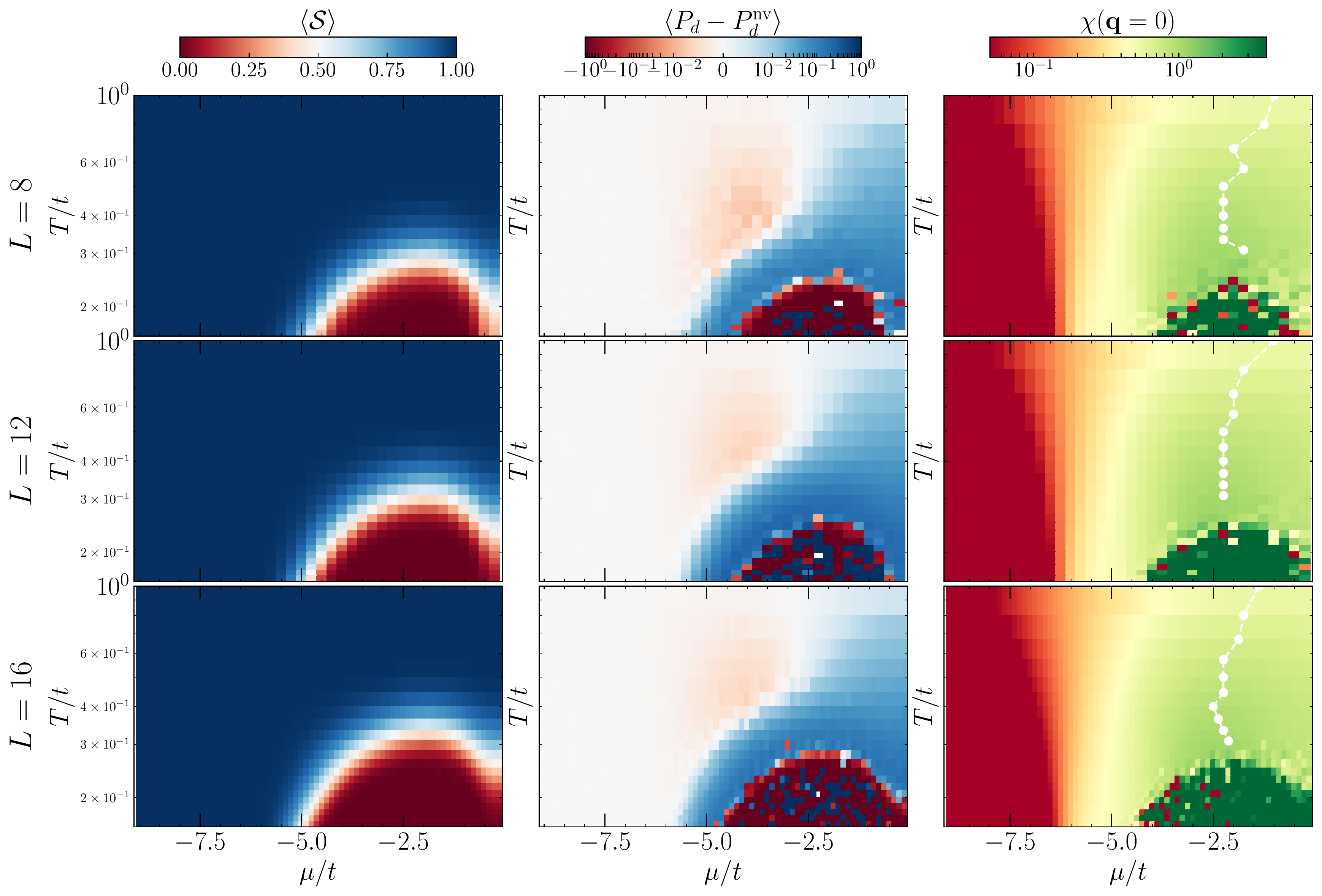}
\caption{\textbf{Finite size effects for the SU(2) Hubbard model on the square lattice.} An analysis of the main quantities and their size dependence for the model with potential connection with the physics of the cuprates. As in the main text, we choose $U/t = 6$ and $t^\prime/t = -0.2$.
%% (\textbf{A}), (\textbf{B}) and (\textbf{C}) 
The top, middle, and bottom rows of plots refer to $L = 8, 12$ and 16, respectively. In turn, the columns from left to right depict the average sign, the correlated $d$-wave pair susceptibility, and the long-wavelength static spin susceptibility. Imaginary-time discretization is fixed at $\Delta\tau = 0.0625$, and all data are extracted as an average of 24 independent runs.}
\label{fig:finite_size_effect_sq_spinful}
\end{figure*}

\paragraph{Finite-size effects.---} An important aspect that deserves further attention concerns finite-size effects on the data presented in the main text. The average sign, for example, when not protected by some underlying symmetry of the Hamiltonian, is known to decrease for larger system sizes~\cite{white89}. Figure~\ref{fig:finite_size_effect_sq_spinful} 
compares the original quantities displayed in the main text [Fig.~\ref{fig:4}] at different lattice sizes.
The ``growth'' of the $\langle {\cal S}\rangle\to 0$ dome is relatively small when taking into account lattices differing by an order of magnitude in the number of sites.  The $d$-wave pair-susceptibility
also qualitatively preserves its overall features, with a tendency of local pair formation at higher electronic densities, encapsulating the $\langle {\cal S}\rangle\to 0$ dome. Similarly, the static spin-susceptibility and its accompanying `pseudogap' line are largely unaltered when the linear lattice size varies from $L=8$ to 16.
These results for different $L$ indicate that our analysis of the link between the SP and quantum criticality is not a finite size effect.

\begin{figure*}[th!]
\includegraphics[width=1.\columnwidth]{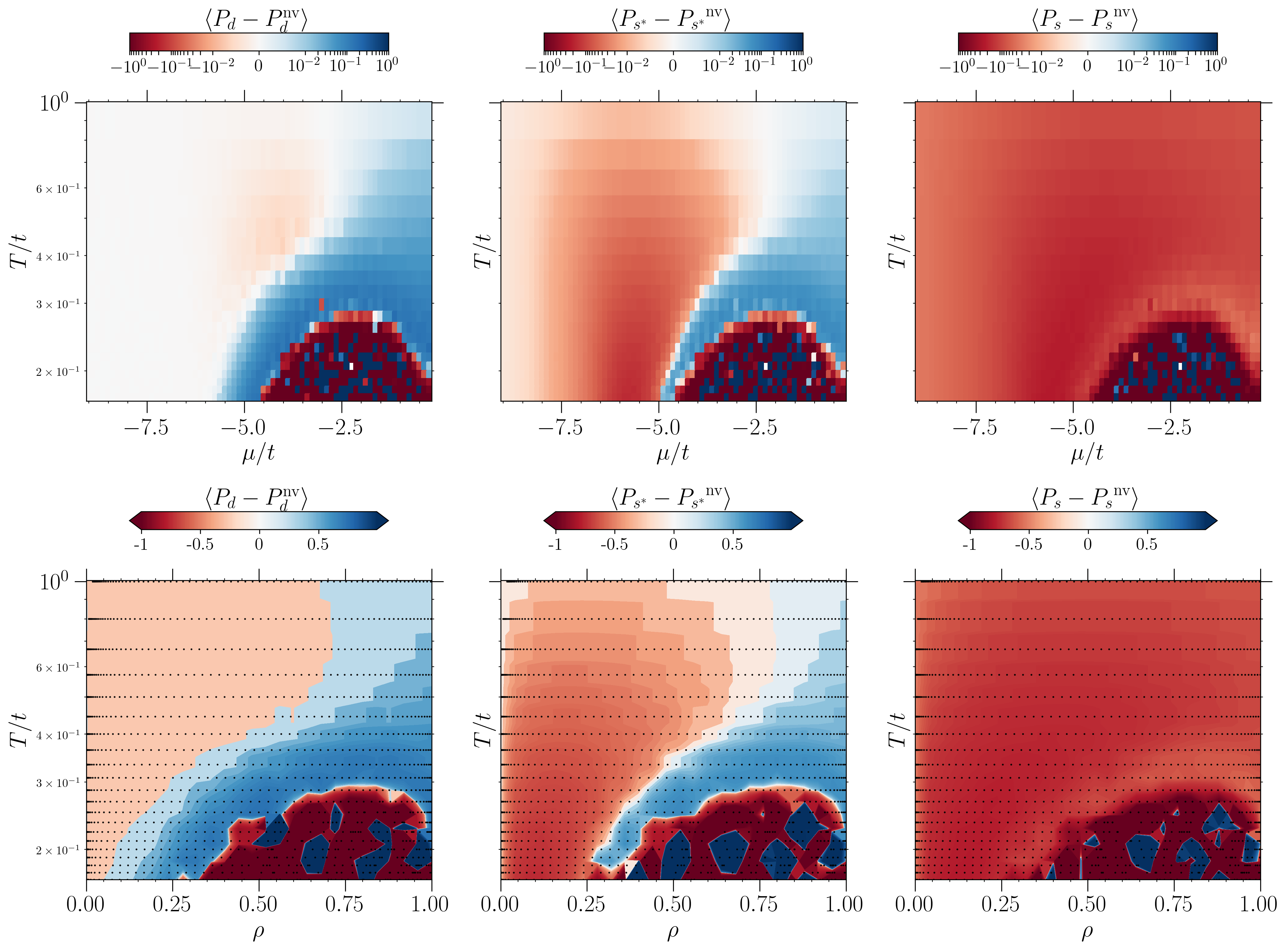}
\caption{\textbf{Comparison of the pairing channels for the SU(2) Hubbard model on the square lattice.} Taking as a starting point for the physics of the cuprates the parameters $U/t = 6$ and $t^\prime/t = -0.2$~\cite{Hirayama2018,Hirayama2019}, we display the comparison of the correlated pair susceptibilities considering different symmetry channels in the left ($d$-wave), center (extended s-wave, $s^*$) and right ($s$-wave) columns, both in $\mu/t$ vs. $T/t$ (upper row) and $\rho$ vs. $T/t$ (lower row) parametric space. Imaginary-time discretization is fixed at $\Delta\tau = 0.0625$, and all data are extracted as an average of 24 independent runs in a lattice with $L=16$.}
\label{fig:s_sx_d_wave_pairing_sq_spinful}
\end{figure*}

\paragraph{Pair-symmetry channels.---} Another important check on the suitability of the chosen parameters to describe the physics of pairs with the same experimentally inferred symmetry as in the cuprates, is to directly compare the correlated susceptibility map with different symmetry channels. This has been done in early studies of QMC~\cite{white89a}, and here, as a side-aspect of the analysis of the average sign, we bring in a systematic investigation. Figure~\ref{fig:s_sx_d_wave_pairing_sq_spinful} displays the correlated pair susceptibility $\langle P_{\alpha} - P_{\alpha}^{\rm nv}\rangle$ dependence on the temperature and electronic filling, with $\alpha = d, s^*$ or $s$-wave. Clearly, the symmetric local pairing channel ($s$-wave), which directly
confronts the on-site $U$, is not favored in the whole range of $T$ and $\rho$ investigated. The correlated pair susceptibility is always negative for finite values of $\langle {\cal S}\rangle$, indicating the vertex is {\it repulsive}. In contrast, both the extended $s$-wave and $d$-wave pairings exhibit positive correlated pairing susceptibilities in the vicinity of the $\langle {\cal S}\rangle\to0$ dome, but are more pronounced in the latter.  This emphasizes the dominance of $d$-wave pairing in the Hubbard model, in direct analogy to a wide class of materials displaying high-temperature superconductivity.

\paragraph{Spectral weight at the anti-nodal point.---}  
In describing the physics of cuprates, a common focus is the anisotropy of the single-particle gap as extracted from ARPES techniques~\cite{Damascelli2003}, which contrasts with the standard isotropic behavior seen in conventional BCS-type superconductors. At the root of the discussion is the shape of the Fermi surface when doping the parent spin-ordered Mott insulator. In particular, to classify the onset of the pseudogap phase at low enough temperatures $T$, i.e., the regime where single-particle, low-energy excitations are suppressed, one tracks the peak of the anti-nodal spectral weight at the Fermi energy, $A_{(\pi,0)}(\omega=0) \equiv -\frac{1}{\pi}{\rm Im}G_{(\pi,0)}(\omega=0)$ as $T$ is varied. 

\begin{figure}[th!]
\includegraphics[width=0.7\columnwidth]{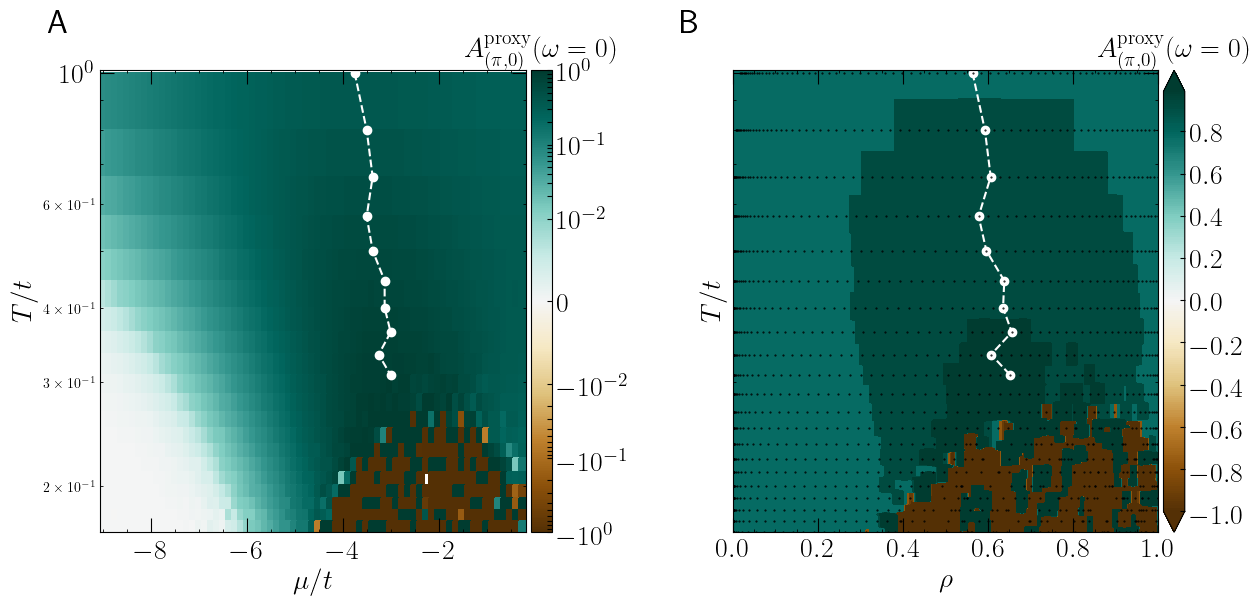}
\caption{\textbf{Comparison of the single-particle spectral weight at the anti-nodal point.} Temperature extrapolation of the extracted $A_{(\pi,0)}$ at the Fermi energy vs the chemical potential $\mu/t$ (\textbf{A}) and the electronic density $\rho$ (\textbf{B}), on a lattice with $L=16$; other parameters are $U/t = 6$ and $t^\prime/t = -0.2$. The maximum values at each temperature are denoted by the white markers. Imaginary-time discretization is fixed at $\Delta\tau = 0.0625$, and all data is extracted as an average of 24 independent runs.}
\label{fig:A_0_sq_spinful_proxy}
\end{figure}

In QMC simulations, this quantity can in principle be extracted by means of an analytical continuation of the data, in which the imaginary-time dependence of the Green's function $G$ is converted to real  frequency. To avoid the well known difficulty of such a calculation~\cite{jarrell96}, a proxy valid at low enough temperatures, $A_{(\pi,0)}^{\rm proxy} = \beta G_{(\pi,0)}(\tau=\beta/2)$ is often used~\cite{trivedi95,Wu2018}. Figure~\ref{fig:A_0_sq_spinful_proxy} shows the results for the anti-nodal spectral function. The pseudogap line extracted from the proxy is qualitatively close to the one obtained from the peak of the static long-wavelength spin susceptibility (Figs.~\ref{fig:4} and  \ref{fig:finite_size_effect_sq_spinful}) and, like it, terminates on  the $\langle {\cal S}\rangle\to 0$ dome.
 
\paragraph{Spectral function and the Lifshitz transition.---} The extensive dataset and associated analysis we have undertaken in this investigation of the sign problem enables us to check other important physical aspects of the Hubbard model on the square lattice, and their relation to cuprate phenomenology.  We include them here, in order to further link their behavior to that of the sign.  One of them refers to the change of the topology of the Fermi surface when increasing the hole-doping (decreasing the electron density) from half-filling: at some critical $\langle n\rangle$, the Fermi surface changes its shape from hole-like to a closed, electron-like one. This transition, referred to as the Lifshitz transition, has been investigated in the context of strongly interacting electrons, and was inferred to
occur concomitantly with the presence of a van Hove singularity at the Fermi level~\cite{Chen2012}. 

\begin{figure}[th!]
\includegraphics[width=0.8\columnwidth]{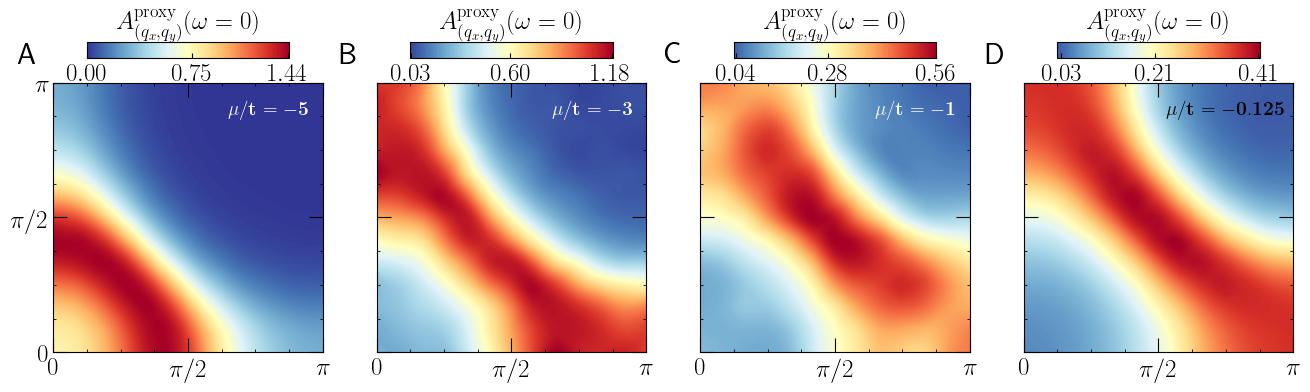}
\caption{\textbf{Lifshitz transition.} The evolution of the spectral function in one quadrant of the Brillouin zone with increasing densities, after an interpolation of the results for a $24\times24$ lattice at temperature $T/t=1/3.25$, with $U/t = 6$ and $t^\prime/t = -0.2$. The different panels depict results with various chemical potentials as marked, and the average density is 0.25 in (\textbf{A}), 0.65 in (\textbf{B}), 0.95 in (\textbf{C}), and 1.00 in (\textbf{D}). Imaginary-time discretization is fixed at $\Delta\tau = 0.0625$, and all data is extracted as an average of 20 independent runs. }
\label{fig:spectral_function}
\end{figure}

Due to the presence of the SP, however, we can only investigate the Lifshitz transition at finite-temperatures, and thus the Fermi `surface' is thermally broadened. Nonetheless, Fig.~\ref{fig:spectral_function} displays the spectral function (obtained via the `proxy' scheme as previously explained) at the Fermi energy on lattices with linear size $L=24$, at a temperature 
%% at which $\langle {\cal S}\rangle$ 
right above the $\langle{\cal S}\rangle\to0$ dome, and investigating dopings above and below the pseudogap line in Figs.~\ref{fig:4} and \ref{fig:spectral_function}. As one increases the density, the change of topology precisely confirms the Lifshitz-scenario, with a further formation of hole-pocket regions along the anti-nodal direction, in direct analogy to the phenomenology of high-Tc superconductors~\cite{Damascelli2003}.

\begin{figure*}[th!]
\includegraphics[width=1.\columnwidth]{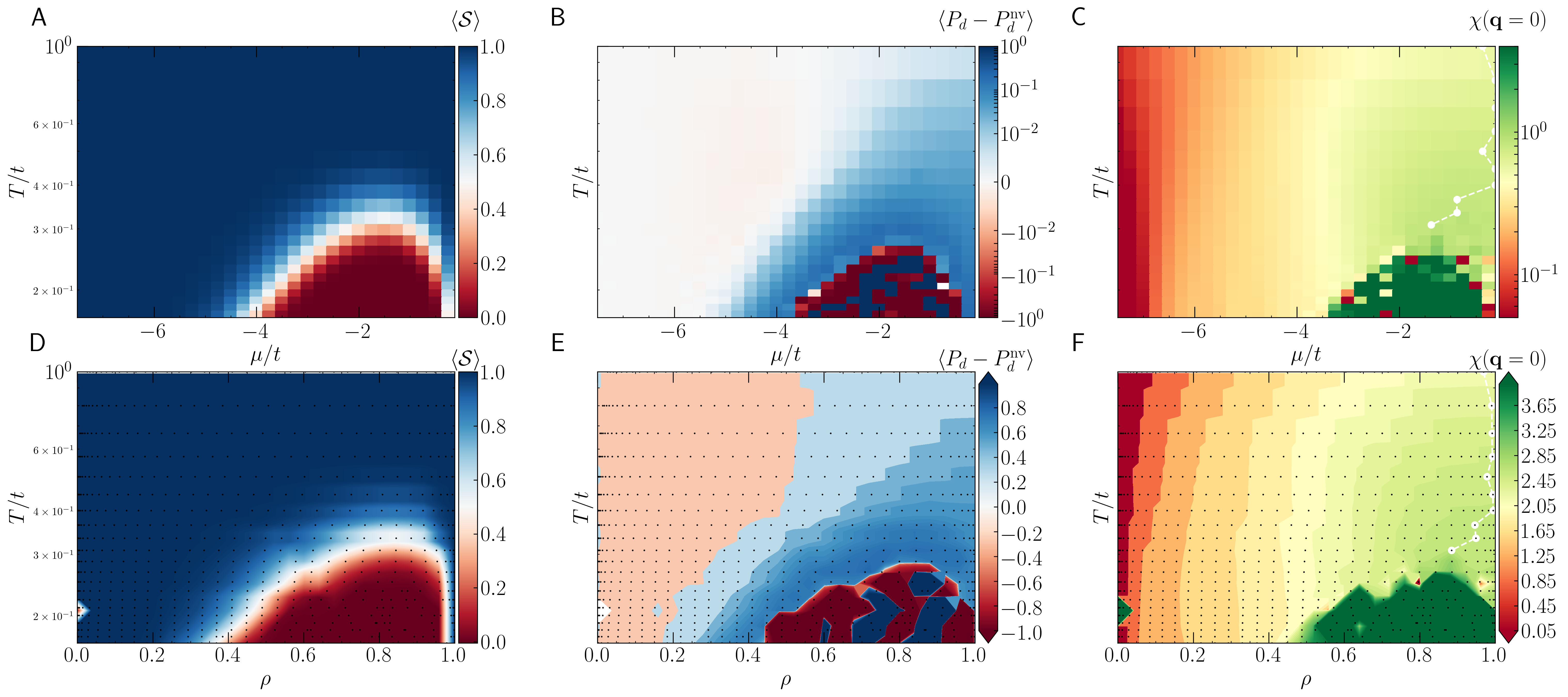}
\caption{\textbf{The `phase diagram' of the bipartite Hubbard model on the square lattice.} The same as Fig.~\ref{fig:4} in the main text, but instead removing the non-bipartite contribution $t^\prime$, i.e., here the `vanilla' Hubbard model results are presented. Other parameters are the same, as $L=16$, $\Delta\tau=0.0625$ and $U/t=6$.}
\label{fig:t2_0_sq_spinful}
\end{figure*}

\paragraph{Comparison to the near-neighbor hopping only Hubbard model.---} 
While it is remarkable how much of the physics of simplest Hubbard Hamiltonian captures that of the  cuprates~\cite{scalapino94}, refinements of the model are known to provide more accurate comparisons to the experiments~\cite{Piazza2012,Hirayama2018,Hirayama2019}. One such, the inclusion of next-nearest neighbor hopping $t^\prime$ was employed in the analysis in the main text.  It is significant in the context of the sign problem because it breaks the particle-hole symmetry and, for example, induces a SP even at half-filling. In this section we address the extent to which including $t^\prime$ affects our conclusions. To this end, we contrast the results of Fig.~\ref{fig:4} with the ones arising from the  Hubbard model with $t'=0$ (Fig.~\ref{fig:t2_0_sq_spinful}). The key qualitative aspects are similar, including the presence of a $\langle {\cal S} \rangle\to0$ dome, the tendency of $d$-wave pair formation (due to the enhanced pair susceptibility with this symmetry around such dome) and a peak of the spin susceptibility ending at the dome. The differences are: (i) while approaching half-filling, the average sign displays a sudden jump towards 1 (as one would expect for this bipartite case), (ii) the extension of the dome, within the temperatures investigated ($T/t \geq 1/6$), is more constrained in density compared to the $t^\prime/t=-0.2$ case, and, more importantly, (iii) the pseudogap region, signified by the temperatures below the $\chi(\textbf{q}=0)$ peak, is significantly reduced.  Our data thus provide another argument in support of an added next-nearest neighbor hopping in order to possibly explain the phase diagram of high $T_c$ materials, which display a robust pseudogap region. However, the main point for the purpose of this manuscript is that whether $t^\prime$ is included or not, the behavior of the sign is correlated with the physics of pairing and magnetism.

\end{document}